\documentclass[%
aip,
amsmath,amssymb,
preprint,%
]{revtex4-1}

\draft 

\usepackage{graphicx}
\usepackage{dcolumn}
\usepackage{bm}

\usepackage[utf8]{inputenc}
\usepackage[T1]{fontenc}
\usepackage{mathptmx}
\usepackage{etoolbox}

\makeatletter
\def\@email#1#2{%
	\endgroup
	\patchcmd{\titleblock@produce}
	{\frontmatter@RRAPformat}
	{\frontmatter@RRAPformat{\produce@RRAP{*#1\href{mailto:#2}{#2}}}\frontmatter@RRAPformat}
	{}{}
}%
\makeatother

\begin{document}

\preprint{https://doi.org/10.1063/5.0212084}
\preprint{AIP/123-QED}

\title[]{Numerical study of synaptic behavior in amorphous HfO$_2$-based ferroelectric-like FETs generated by voltage-driven ion migration} 



\author{J. Cuesta-Lopez}
\author{M.D. Ganeriwala}
\author{E.G. Marin}
\affiliation{Departamento de Electrónica y Tecnología de Computadores, Universidad de Granada, Granada 18071, Spain}
\author{A. Toral-Lopez}
\affiliation{Dipartimento di Ingegneria dell'Informazione, Università di Pisa, 56122 Pisa, Italy}
\author{F. Pasadas}
\author{F.G. Ruiz}
\author{A. Godoy}
\affiliation{Departamento de Electrónica y Tecnología de Computadores, Universidad de Granada, Granada 18071, Spain}
\email{agodoy@ugr.es}
\email{jcuesta@ugr.es}


\date{\today}

\begin{abstract}
The continuous effort in making artificial neural networks more alike to human brain calls for the hardware elements to implement biological synapse-like functionalities. The recent experimental demonstration of ferroelectric-like FETs promises low-power operation as compared to the conventional ferroelectric switching devices. This work presents an \textit{in-house} numerical tool, which self-consistently solves the electrostatics and time-dependent electronic and ionic transport. The tool is exploited to analyze the effect that various physical parameters such as mobility and ion concentration could have on the design of the ferroelectric-like FETs. Their suitability in emulating different functions of the biological synapses is also demonstrated. 
\end{abstract}

\pacs{}

\maketitle 

\section{Introduction}
The nascent era of Artificial Intelligence (AI) demands the processing and storage of huge amounts of data, which, from the point of view of energetic efficiency, can hardly be supported by traditional von-Neumann architectures \cite{Roy2019,Najafabadi2015}. Consequently, there is a strong need for finding new physical implementations of artificial neural networks (ANNs), more alike to the human brain processing schemes \cite{Mead1989,Zhu2020}. The unit elements of such ANNs should resemble the operation of the biological synapses, characterized by an hysteretic behavior with ability to alter its strength in response to applied stimuli, the so-called plasticity,\cite{Hebb1949,Indiveri2015,Zidan2018}. Different device architectures have been proposed to this purpose, including prevailing resistive random access memories (RRAMs) based on the formation of conductive filaments \cite{Ielmini2011,Larentis2012,Wang2023}, as well as less conventional options using polymeric electrolyte based devices \cite{Zhu2018}, redox \cite{Fuller2019} or ferroelectric transistors \cite{Jerry2017,Si2019,Chen2020,Zhu2018}. 

Among these competitors to mainstream RRAMs, Ferroelectric Field-Effect Transistors (FeFETs) have already exhibited lower power consumption and faster operation \cite{Kim2021}, as well as a high endurance and retention time \cite{Chatterjee2017,Ni2018}. The memory and logic functions of FeFETs can be physically implemented in distinct ways. Most novel FeFETs adjust to a 1-transistor (1T) in-memory computing structure, with the ferroelectricity built-in the semiconductor channel \cite{Osada19}. However, charge trapping at the semiconductor interface and leakage currents result in short retention times, a drawback that needs to be addressed before reaching the market \cite{Si2019}. Nevertheless, the most common structure resembles the traditional 1-transistor-1-capacitor (1T1C) architecture, where the ferroelectric effect is provided by a dielectric oxide with pollycrystalline structure, such as HfO$_2$ or ZrO$_2$ \cite{Muller2013,Liu2019}. Unfortunately, FeFETs based on nanocrystal ZrO$_2$ are not compatible with current CMOS technology \cite{Peng2020b}, whilst pollycrystalline doped-HfO$_2$-based FeFETs are prone to suffer from high-power consumption due to undesirable leakage currents and require annealing processes up to $500$ ºC to form orthorhombic crystal phases \cite{Peng2020}.

A potential alternative to the material stack of FeFET (while keeping the principle of operation) is represented by ferroelectric-like FETs, which make use of the memory behavior arising from voltage-driven oxygen vacancy and negative ion migration in amorphous oxides \cite{Lim2023}. The resulting switchable polarization states of the oxide (i.e. the alternate migrations of anions and cations across the oxide thickness), can be induced by applying voltage pulses to the gate contact, originating the consequent modulation of the channel conductance \cite{Liu2022,Zhang2021}. This phenomenon, which has been confirmed in experimental realizations \cite{Feng2021}, enables the emulation of input pulse dependent synaptic functions \cite{Mul2017}, making ferroelectric-like FETs suitable for the implementation of Spiking Neural Networks (SNNs) \cite{Roy2019}. Moreover, devices based on amorphous dielectrics, such as amorphous HfO$_2$ or Al$_2$O$_3$ \cite{Peng2019}, show a better compatibility with CMOS technology than polycrystalline HfO$_2$-based FeFETs \cite{Peng2021}, and require a lower operation voltage \cite{Peng2022}, similar to the action potentials of bio-synapses.

The understanding of the physical mechanism controlling the kinetics of the involved ions is paramount to the future progress and eventual adoption of ferroelectric-like technologies in SNN. In spite of some experimental \cite{Peng2020b, Lim2023}, and exploratory numerical studies \cite{Zheng2021,Chen2022,Feng2021,Wang2015,Kumar2018}, there is still a strong need of deeper comprehension so to unravel the optimal design and operation of these devices. This work presents a numerical study of the physical phenomena and operating principles in ferroelectric-like FETs driven by the redistribution of ions inside amorphous oxides. More precisely, starting by an experimental validation that accounts for the suitability of the \textit{in-house} numerical tool, we assess the impact of the oxide-ion mobility and concentration, as well as of the input pulses in the response of a HfO$_2$/Ge-based ferroelectric-like FET. The dependence of various electrical figures-of-merit on the physical parameters of the FET are analyzed to elucidate their optimal configuration.

\section{Methods}\label{sec:methods}

\begin{figure}[!t]
	\begin{center}
		\includegraphics[width=0.45\textwidth]{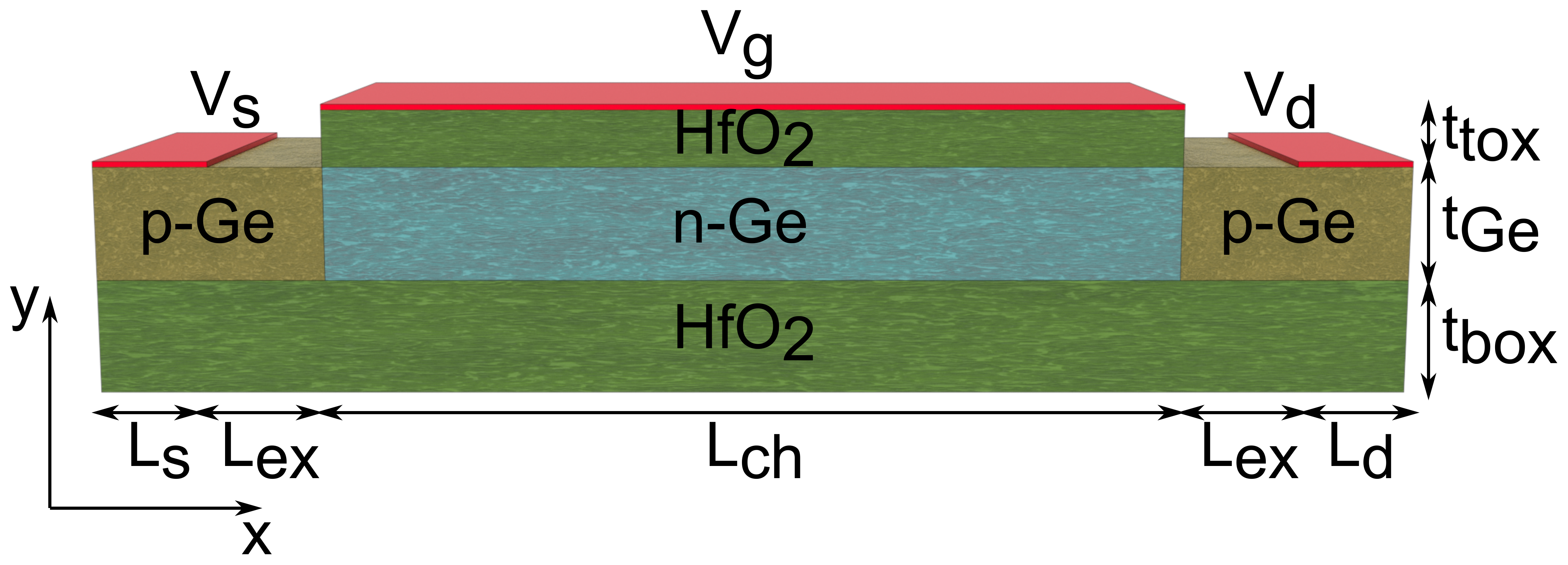}
		\caption{\label{fig:device} Schematic of the simulated device. Here, an n-type Ge channel of thickness t\textsubscript{Ge} is sandwich between top amorphous and bottom HfO$_2$ oxide layers of thicknesses t\textsubscript{tox} and t\textsubscript{box}, respectively. The channel length is denoted by L\textsubscript{ch}, while the source/drain length and extension region by L\textsubscript{s/d} and L\textsubscript{ex}, respectively.}
	\end{center}
\end{figure}

The device considered in the study is schematically depicted in Figure \ref{fig:device}. A Germanium channel with thickness t\textsubscript{Ge} is sandwiched between two HfO$_2$ regions, with thicknesses t\textsubscript{tox} (top amorphous layer) and t\textsubscript{box} (bottom layer). The channel length is denoted by L\textsubscript{ch} and the source and drain regions lengths are L\textsubscript{s} and L\textsubscript{d}, respectively. A source/drain extension region of length L\textsubscript{ex} is also included.

Simulations are performed employing an \textit{in-house} developed numerical tool that self-consistently solves the electrostatics and time-dependent electronic and ionic transports under a semi-classical scheme. The equation system comprises the Poisson equation (eq. \ref{eq:Poisson}), the time-dependent continuity equation for electrons and holes based on the pseudo-Fermi energy level (eq. \ref{eq:conteh}) \cite{Toral-Lopez2022}, solved in the Ge channel, and the time-dependent continuity equation for ions in the amorphous HfO$_2$, following a drift-diffusion scheme after applying the Scharfetter-Gummel method (eq. \ref{eq:conti}) \cite{SG1969}.
\begin{equation}\label{eq:Poisson}
	\vec{\nabla}\cdot\left(\varepsilon \vec{\nabla} V\right) = - \rho
\end{equation}
\begin{equation}\label{eq:conteh}
	\left\{\begin{array}{l}
		\vec{\nabla} \cdot \vec{J_{\rm n}} = \vec{\nabla} \cdot \left[ q \mu_{\rm n} n \vec{\nabla} E_{\rm F, n} \right] = + q \frac{\partial n}{\partial t}\\
		\vec{\nabla} \cdot \vec{J_{\rm p}} = \vec{\nabla} \cdot \left[ q \mu_{\rm p} p \vec{\nabla} E_{\rm F, p} \right] = - q \frac{\partial p}{\partial t}
	\end{array}
	\right.
\end{equation}
\begin{equation}\label{eq:conti}
	\vec{\nabla} \cdot \vec{J_{\rm i}} = \vec{\nabla} \cdot \left[ - z_{\rm i} q D_{\rm i} e^{-s_{\rm i} \Phi} \vec{\nabla} \left( c_{\rm i} e^{s_{\rm i} \Phi} \right) \right] = - z_{\rm i} q \frac{\partial c_{\rm i}}{\partial t}
\end{equation}
Here, $V$ is the electrostatic potential, $\rho$ is the charge density, $\varepsilon$ is the dielectric constant, $\vec{J}$ is the current density, $q$ is the elementary charge unit, $\mu$ is the mobility, $n$ and $p$ are the electron and hole densities respectively, $E_{\rm F}$ is the quasi-Fermi level, $t$ is the time, $z$ is the ion valence with $s=\left|z\right|/z$ the ion valence sign, $c$ is the ion concentration, $D$ is the diffusion coefficient, which follows the Einstein relationship ($D = \mu k_{\rm B} T/q$), $k_{\rm B}$ is the Boltzmann constant, $T$ is the temperature, $\Phi = q V/(k_{\rm B} T)$ is the normalized potential, and the sub-indexes $n$, $p$ and $i$ denote electrons, holes and each ion specie, respectively.

Amorphous HfO$_2$ is known to have a high intrinsic concentration of negatively charged oxygen ions (O\textsuperscript{2-}) and positively charged oxygen vacancies (V\textsubscript{O}\textsuperscript{2+}) \cite{Foster2002}, both of which are considered as mobile charges in the simulator. Note that both ions and vacancies are confined inside the top amorphous HfO$_2$ layer, i.e. they can drift and diffuse inside that material region, but are not allowed to penetrate the interfaces. The electron and hole concentrations in the channel are evaluated using the density of states, g(E), and the Fermi-Dirac distribution function: 
\begin{equation}\label{eq:eh}
	\left\{\begin{array}{l}
		n = \int_{E_{\rm c}}^{+\infty}g(E)f(E)dE\\
		p = \int_{-\infty}^{E_{\rm v}}g(E)\left[1-f(E)\right]dE\\
	\end{array}
	\right.
\end{equation}
where E$_{\rm c}$ and E$_{\rm v}$ are the energies corresponding to the bottom of the conduction band and the top of the valence band, respectively. The numerical tool also includes the modelling of interface traps according to the details given in the Appendix.


\section{Experimental validation}\label{sec:exp}

In order to validate the numerical simulator, its results are compared with the experimental data presented in \cite{Peng2022}. The fabricated device consists of a Ge-based p-type MOSFET with L\textsubscript{ch} = 3 \textmu m and an amorphous HfO$_2$ gate insulator with t\textsubscript{tox} = 3 nm, exhibiting a noticeable synaptic behavior. Acceptor and donor doping of N$_\text{A} = 10^{17}$ cm$^{-3}$ and N$_\text{D} = 10^{12}$ cm$^{-3}$ are set in the source/drain and channel regions, respectively. The theoretical mobility of electrons and holes in the Ge channel is scaled by a factor $1/6$, the same used to reduce the channel length from L\textsubscript{ch} = 3 \textmu m to L\textsubscript{ch} = 500 nm in the simulated structure, thus minimizing the computational burden. Then, the mobility was adjusted to match the experimental value of the saturation current, $\mu$\textsubscript{n} = 144 cm$^2$/Vs and $\mu$\textsubscript{p} = 70 cm$^2$/Vs. The value of the dielectric constant of Ge and HfO$_2$ are fixed to $\varepsilon$\textsubscript{Ge} = 16$\varepsilon_0$ \cite{Sze2012} and $\varepsilon$\textsubscript{HfO\textsubscript{2}} = 20$\varepsilon_0$ \cite{Ceresoli2006}, respectively. The metallic contacts are modeled as ohmic, and the difference between their metal workfunction, $\phi_{\rm m}$, and the semiconductor electron affinity, $\chi_{\rm sc}$, is determined by shifting the threshold voltage of the transfer characteristics, V\textsubscript{T}, obtaining a value of $\chi_{\rm sc} - \phi_{\rm m} = -0.206$ eV. The values of L\textsubscript{s/d} and L\textsubscript{ex} are set to $20$ nm and $40$ nm, respectively. 

\begin{figure}[!t]
	\begin{center}
		\includegraphics[width=1\linewidth]{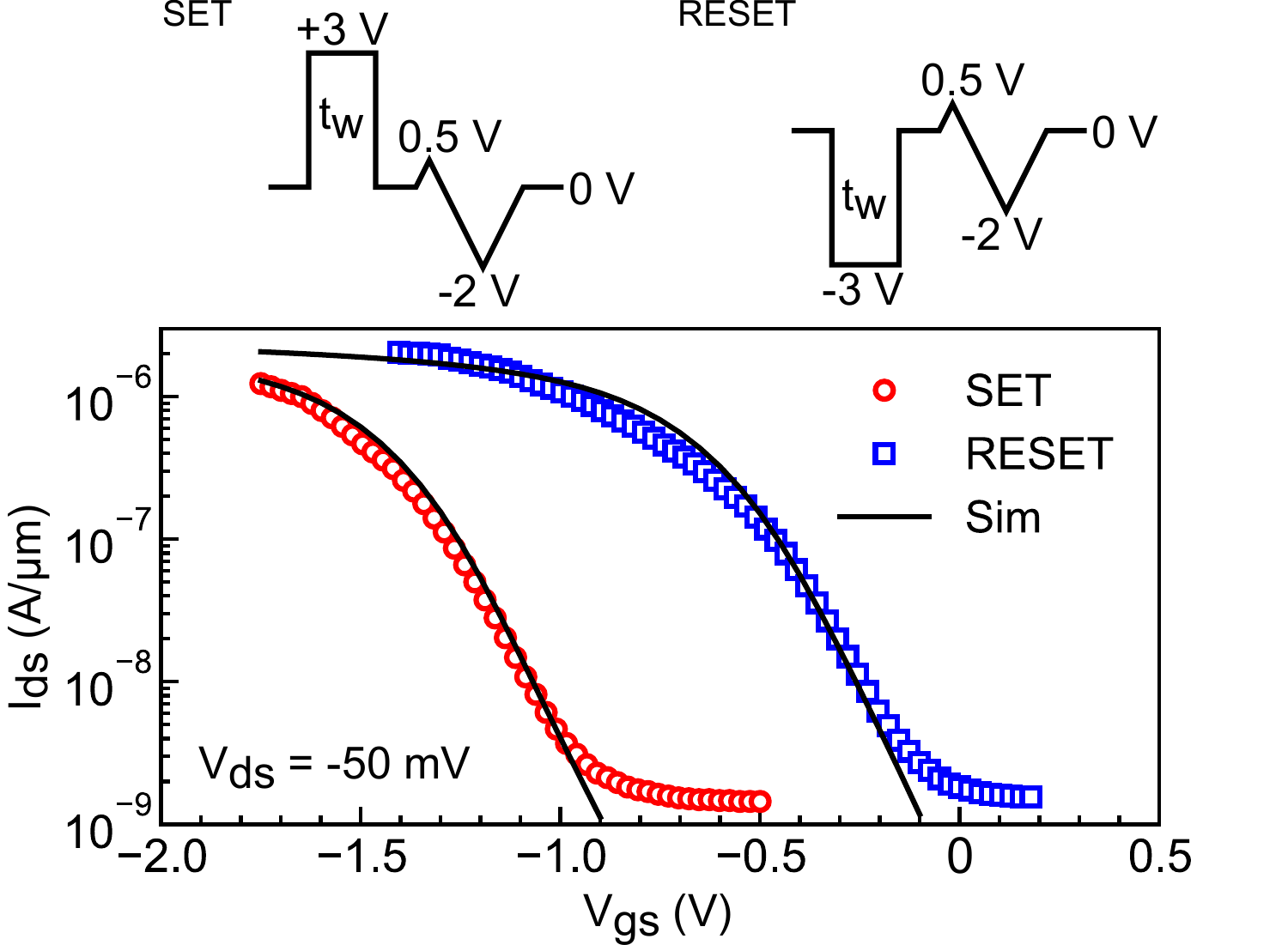}
		\caption{\label{fig:expComp} Comparison between simulated data (solid lines) and experimental data (symbols) from \cite{Peng2022}, showing a good agreement. The SET/RESET pulses are of amplitude $\pm$3 V and width t\textsubscript{w} = 1\textmu s at V\textsubscript{ds} = 0 V, while the current is read at V\textsubscript{ds} = -50 mV.}
	\end{center}
\end{figure} 

The transfer characteristics (I\textsubscript{ds}-V\textsubscript{gs}) depicted in Figure \ref{fig:expComp} show a very good agreement between simulations and experimental data, validating the suitability of the simulator to perform the theoretical study. To achieve these results, a square pulse is first applied at the gate terminal with amplitude V\textsubscript{gs} = $\pm$3 V and width t\textsubscript{w} = 1 \textmu s, at a fixed drain voltage V\textsubscript{ds} = 0 V. The application of a positive and negative gate pulse, respectively, brings the device to a high-V\textsubscript{T} state (SET) or to a low-V\textsubscript{T} state (RESET), caused by the redistribution of ions within the amorphous gate oxide. The SET/RESET process is followed by a linear sweep of V\textsubscript{gs}, with a lower amplitude, enough to read the changed conductivity of the FET without perturbing the ion distribution. During the read operation, V\textsubscript{ds} is set to -50 mV. The same biasing scheme and voltages are henceforth used unless otherwise mentioned explicitly.

The values \textmu$_{\rm i} = 10^{-9}$ cm$^2$/(Vs), and c$_{\rm i} = 2.5\times10^{20}$ cm$^{-3}$, are found by fitting the experimental results. Initially, the ion concentration is uniformly distributed in the 3-nm thick HfO\textsubscript{2}. Some experimental works have reported that a higher ion density can be measured at the metal/oxide interface than at the oxide/semiconductor interface \cite{Peng2020,Wang2015,Feng2021}. Nevertheless, given the thin insulating layer and the operating frequencies of the applied signals, this initial scenario would have negligible effect on the later self-consistent redistribution of ions across the insulating layer. Regarding the ions mobility, it has been proven that the mobility of positively charged oxygen vacancies is lower than that of the negative oxygen ions \cite{Zafar2010}, hence why some authors consider oxygen vacancies to be stationary \cite{Chen2022}. Nonetheless, several simulation works ignore this difference between ion mobilities due to its negligible effect on the results \cite{Lim2023,Marchewka2016,Kumar2018}. Thus, we proceeded under the latter assumption.

The Ge density of states, g(E), is computed making use of the density functional theory (DFT) (see Figure \ref{fig:Ge} in Appendix). Additionally, some works have shown that the trapping/de-trapping of carriers at the interfaces could be responsible of the clockwise hysteretic behavior when operating at frequencies up to f $\approx$ 1 kHz \cite{Daus2017}. This phenomenon can be considered by including the traps at the HfO$_2$/Ge interface (see Figure \ref{fig:traps} in the Appendix). However, since the input signal here is in the MHz range \cite{Peng2022}, the traps are assumed to be stationary.


\section{Results and discussion}

Next, the operation of ferroelectric-like FETs is investigated in detail, to predict their potential performance when integrated in ANNs. The geometrical parameters henceforth used for device simulations are listed in Table \ref{tab:para}. The rest of the physical parameters used for the experimental validation remain unchanged, except for the electron and hole mobility which are now set to their conventional values, $\mu_{\rm n} = 864$ cm$^2$/(Vs) and $\mu_{\rm p} = 420$ cm$^2$/(Vs). We also assume to discard the presence of traps due to their small impact in the simulations as they do not contribute to hysteresis, but only to small adjustments on V\textsubscript{T}.

\begin{table}[!t]
	\caption{\label{tab:para} Geometrical parameters of the simulated device (Figure 1) used for the analyses.}
	\begin{ruledtabular}
		\begin{tabular}{cccccccc}
			Parameter & L\textsubscript{ch} & L\textsubscript{s} & L\textsubscript{d} & L\textsubscript{ex} & t\textsubscript{Ge} & t\textsubscript{tox} & t\textsubscript{box} \\
			\hline
			Value  (nm)   & 150   & 20   & 20   & 20   &  20   &  3    &  20 \\
		\end{tabular}
	\end{ruledtabular}
\end{table}

\begin{figure*}[]
	\begin{center}
		\includegraphics[width=1\linewidth]{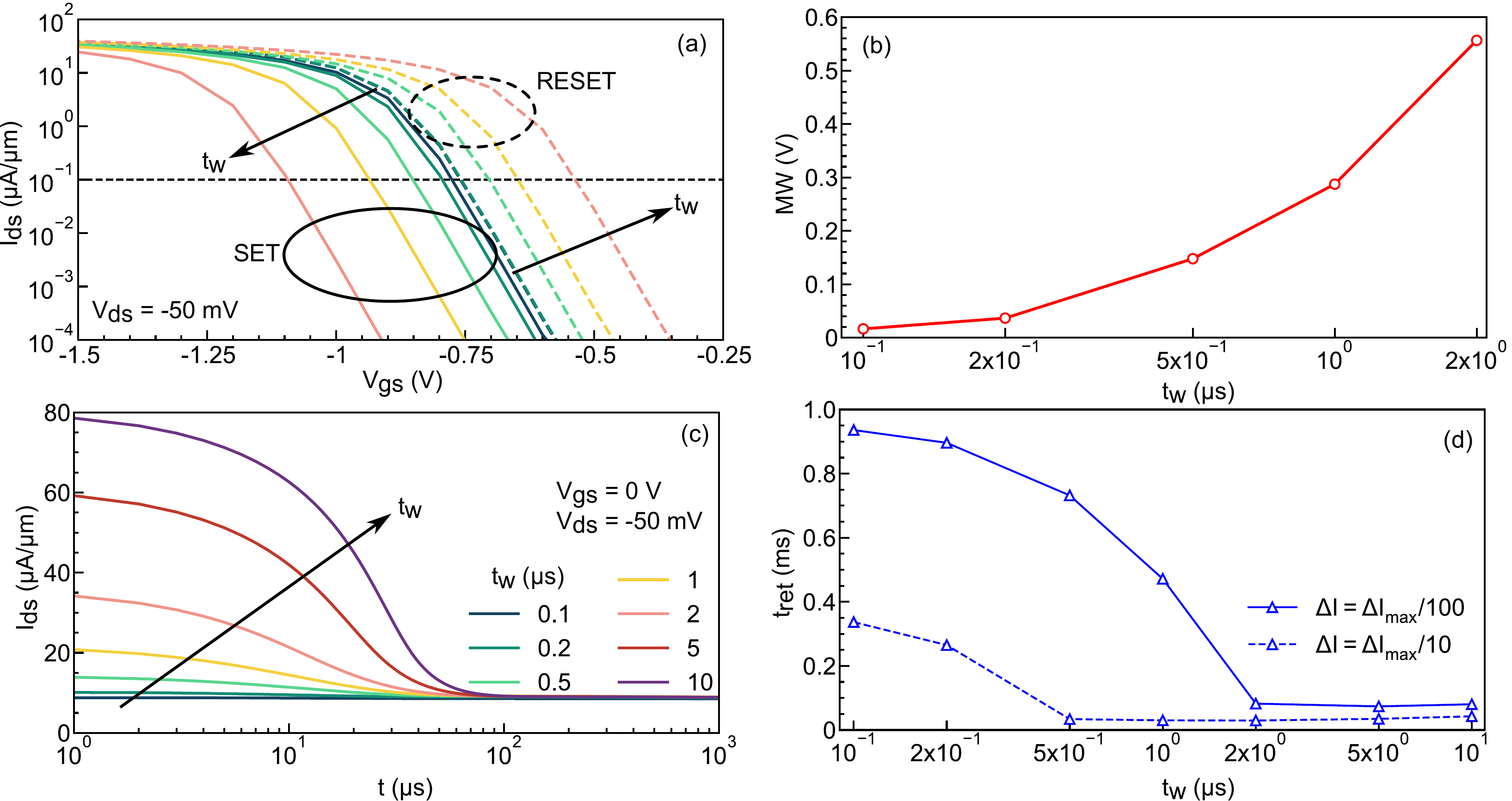}
		\caption{\label{fig:tw} \textbf{(a)} Transfer characteristics for different SET/RESET pulse widths, t$_{\rm w}$. Solid and dashed lines represent I\textsubscript{ds} after the SET and the RESET pulse, respectively. \textbf{(b)} Memory window defined as MW $=$ V$_{\rm T}$(RESET) $-$ V$_{\rm T}$(SET) as a function of t\textsubscript{w}. \textbf{(c)} Time evolution of I\textsubscript{ds} after the end of the SET pulse for different t\textsubscript{w}, showing a gradual decrease to its steady state value. \textbf{(d)} Retention time (t$_{\rm ret}$), defined for $\Delta$I = $\Delta$I\textsubscript{max}/100 (solid) or $\Delta$I = $\Delta$I\textsubscript{max}/10 (dashed), as a function of t\textsubscript{w}.}
  \end{center}
\end{figure*}

\begin{figure*}[]
	\begin{center}
		\includegraphics[width=\textwidth]{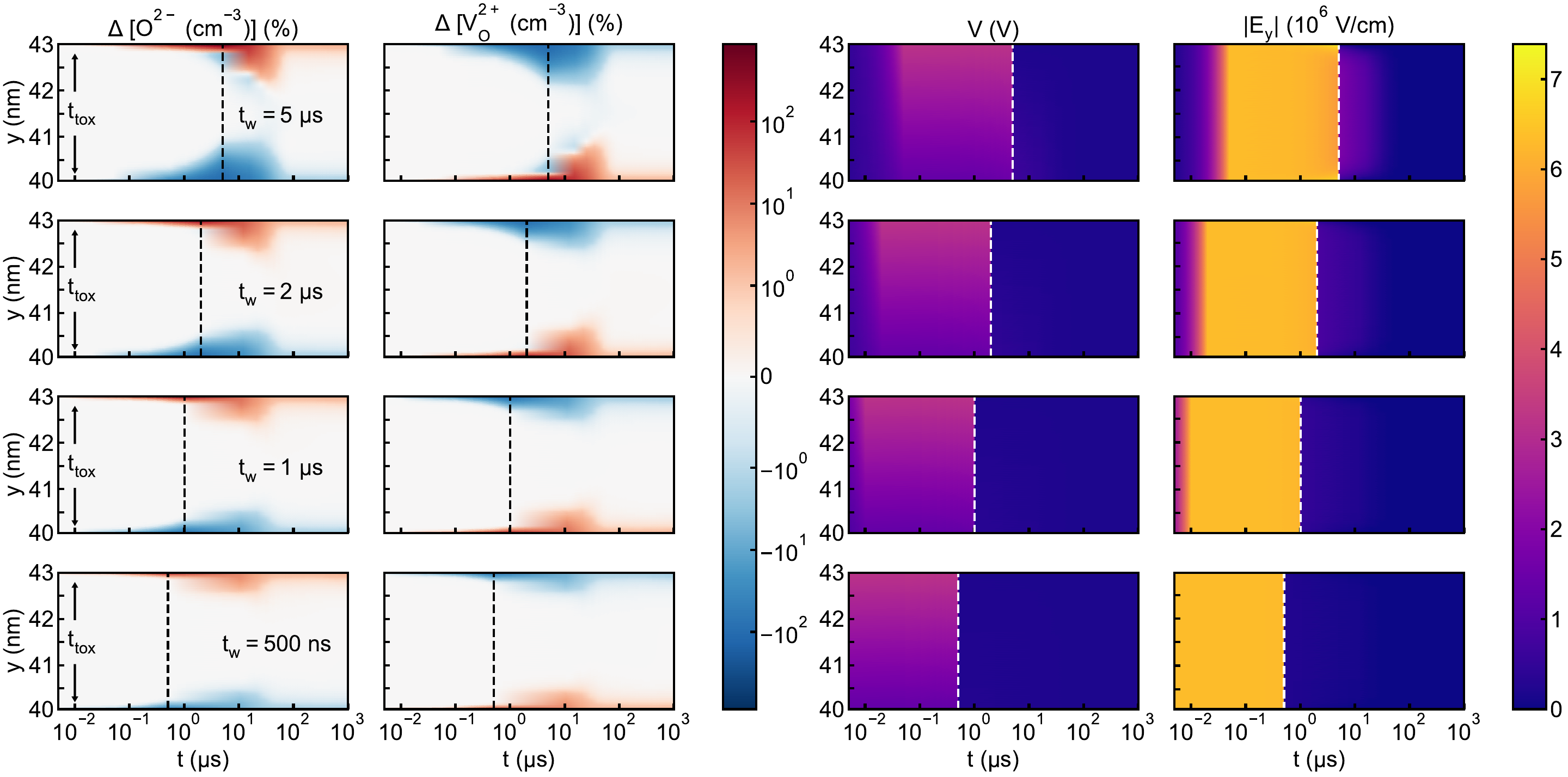}
		\caption{\label{fig:off} Colormaps depict, from left to right, the relative variation of the concentration of oxygen ions (O\textsuperscript{2-}) and vacancies (V\textsubscript{O}\textsuperscript{2+}) with respect to c$_{\rm i}$ in logarithmic scale, the electrostatic potential and the vertical electrostatic field, measured across the amorphous HfO$_2$ layer thickness at several time instants from t = 0 s to t = t$_{\rm w}$ + 1 ms, where t$_{\rm w}$ is the duration of the SET pulse. Data is represented, from top to bottom, for t$_{\rm w} = 0.5$, $1$, $2$ and $5$ \textmu s. Vertical dashed lines delimit the duration of the SET pulse.}
	\end{center}
\end{figure*}

Figure \ref{fig:tw}a shows the transfer characteristics of the ferroelectric-like FET upon application of a fixed amplitude SET/RESET pulse with V\textsubscript{gs} = $\pm$3 V and varying t\textsubscript{w}. The progressive change in V\textsubscript{T}, achieved by modifying the t\textsubscript{w}, can be used to store multibit binary data or modulate the weight change in synaptic emulation. The maximum change in the conductivity obtained upon SET/RESET of the device can be captured by the memory window (MW), defined as MW $=$ V$_{\rm T}$(RESET) $-$ V$_{\rm T}$(SET). Note that the V\textsubscript{T} of the device is calculated at a constant current of 0.1 \textmu A/\textmu m. Fig. \ref{fig:tw}b proves that a larger t\textsubscript{w} increases the resulting MW. 

Shortly after the end of the SET/RESET pulse, the ions start diffusing back, giving the ferroelectric-like FET its volatile behavior. Figure \ref{fig:tw}c shows the I\textsubscript{ds} time evolution after the end of a SET pulse. As shown, the current I\textsubscript{ds} gradually decreases to its steady state value due to ions diffusing back to their equilibrium position. The retention time (t\textsubscript{ret}) is defined as the time required for
the drain current to drop to a 1$\%$ of its maximum value \cite{Farronato2022}, as follows: t$_{\rm ret} = \Delta$t @ $\Delta$I = $\Delta$I\textsubscript{max}/100. Here, the maximum drain current is the one measured right after the writing pulse of duration t\textsubscript{w}, I\textsubscript{ds}(t\textsubscript{w}), and currents are referenced to the steady state value, I\textsubscript{ds}(st.), so that $\Delta$I\textsubscript{max} = $\text{I}_{\rm ds}(\text{t}_{\rm w}) - \text{I}_{\rm ds}(\text{st.})$ and $\Delta$I(t) = $\text{I}(\text{t}) - \text{I}_{\rm ds}(\text{st.})$. It is noteworthy that the value of t\textsubscript{ret}, plotted in Fig. \ref{fig:tw}d, decreases with increasing t\textsubscript{w}, a behavior opposite to that of MW. An alternative definition of t\textsubscript{ret} that considers the current dropping to a 10$\%$ of $\Delta$I\textsubscript{max} is also analyzed (Figure \ref{fig:tw}d, dashed lines), showing the same monotonically decreasing behavior of t\textsubscript{ret}.

\begin{figure*}[t]
	\begin{center}
		\includegraphics[width=\textwidth]{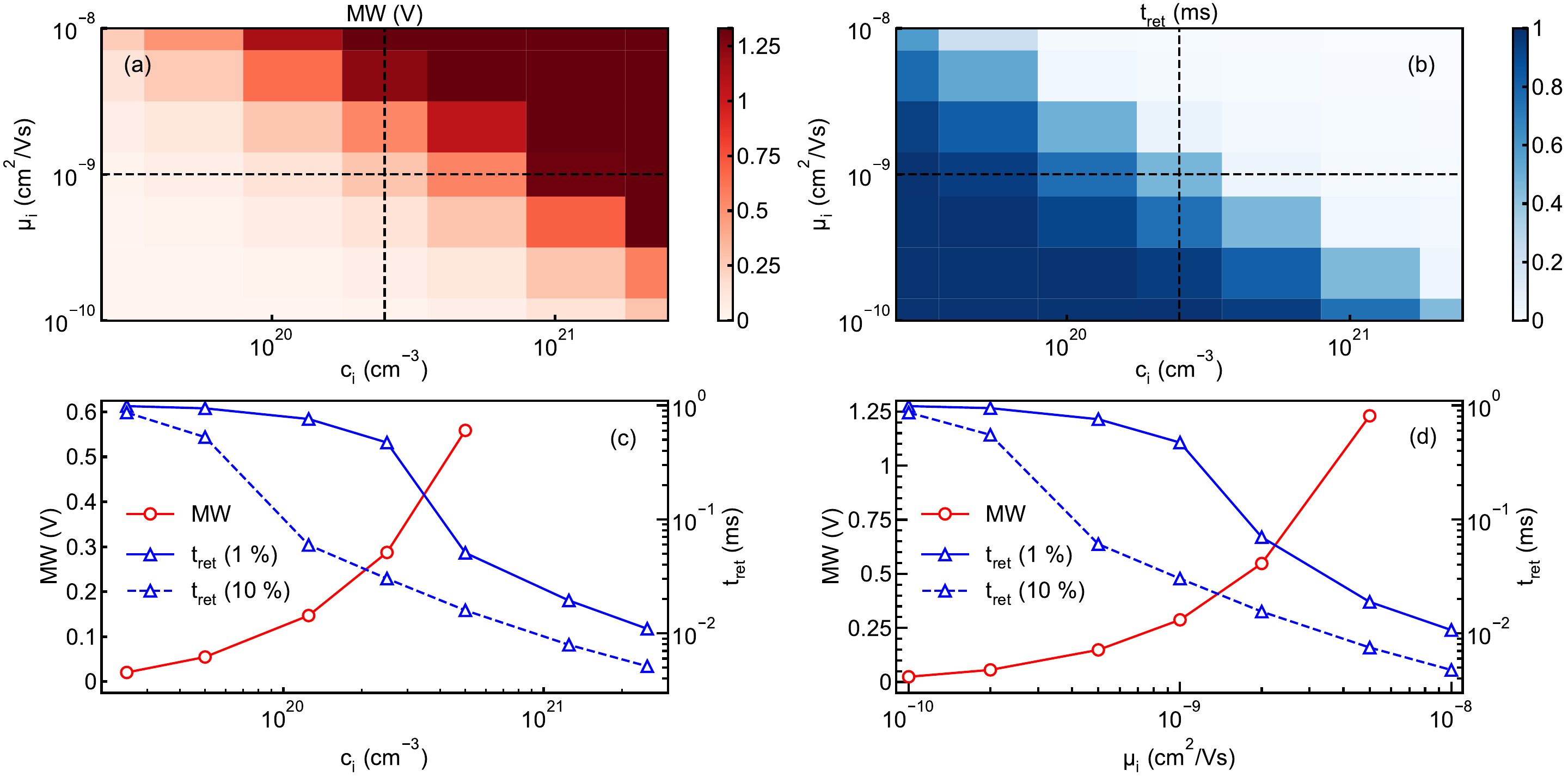}
		\caption{\label{fig:trade-off} (a) Memory Window (MW) and (b) Retention time (t\textsubscript{ret}) as a function of the ion concentration (c\textsubscript{i}) and its mobility (\textmu \textsubscript{i}). Trade-off between the MW and t\textsubscript{ret} with respect to (c) c\textsubscript{i} and (d) \textmu\textsubscript{i}. The retention time was defined as the time required for I\textsubscript{ds} to drop to a 1$\%$ (solid) or a 10$\%$ (dashed) of the maximum initial current.}
	\end{center}
\end{figure*}

To gain further insights into the opposing behavior of MW and t\textsubscript{ret}, let us consider Figure \ref{fig:off}, which shows the oxygen ions and vacancies distribution, as well as the potential and electric field across the gate oxide thickness as a function of time and for different t\textsubscript{w}. A SET pulse of V\textsubscript{gs} = $3$ V at V\textsubscript{ds} = 0 V is used and each row stands for a t\textsubscript{w} value, from 500 ns to 5 \textmu s. As can be seen, both ions and vacancies are drifted by the SET pulses from the initial uniform distribution towards the interfaces of the oxide layer, with oxygen ions accumulating under the metallic contact and vacancies moving towards the channel. This redistribution, more accentuated for longer t\textsubscript{w} values, gives rise to a higher electric field (E-field) as well. The peak absolute value of the E-field across the gate oxide is $\approx$ 6 MV/cm and $\approx$ 7 MV/cm for t\textsubscript{w} = 500 ns and 5 \textmu s, respectively. Therefore, with increasing t\textsubscript{w}, the internal E-field (in direction opposite to the externally applied E-field) also increases, further shifting the channel V\textsubscript{T} and resulting in a wider MW. When the stimulus disappears (dashed vertical line, V\textsubscript{gs} = 0 V after t\textsubscript{w}), ions continue drifting. Their low mobility causes this delayed reaction and, as a consequence, the hysteretic behavior of ferroelectric-like FETs. When the drift ends, ions start diffusing back from the interfaces, eventually reaching the initial uniform distribution. It can be seen that the longer the duration of the pulse, the higher the concentration of ions at the interfaces. The maximum concentration of oxygen ions at the interfaces are found to be $\approx$ 3.9x10\textsuperscript{20} cm\textsuperscript{-3} and $\approx$ 2.43x10\textsuperscript{21} cm\textsuperscript{-3} for t\textsubscript{w} = 500 ns and 5 \textmu s respectively. In conclusion, the longer the t\textsubscript{w}, the higher the drift and the built-in E-field that gives rise to a wider MW. However, at the same time, a higher t\textsubscript{w} increases the concentration gradient, therefore increasing the diffusion and reducing the t\textsubscript{ret}, giving rise to the aforementioned trade-off between MW and t\textsubscript{ret} with respect to t\textsubscript{w}.

The drift and diffusion of ions, and specially their influence on MW and t\textsubscript{ret}, is expected to depend on the mobility (\textmu$_{\rm i}$) and the concentration (c$_{\rm i}$) of ions. Beside being appropriate for its use as synaptic element, depending on the values of the MW and t\textsubscript{ret}, the ferroelectric-like FET could find varied applications, such as memory selector, true random number generator or in artificial neurons \cite{Chekol2022}. Therefore, the better understanding of the dependence of MW and t\textsubscript{ret} on \textmu$_{\rm i}$ and c$_{\rm i}$ could provide accurate guiding principles to optimize the design of ferroelectric-like FETs for targeted applications. 
Figure \ref{fig:trade-off}a shows the MW as a function of \textmu$_{\rm i}$ and c$_{\rm i}$ for t\textsubscript{w} = 1 \textmu s. It can be seen that higher values of \textmu$_{\rm i}$ (resulting into higher drift velocities) and c$_{\rm i}$ (producing larger gradients of ion concentrations) enable larger MWs as larger E-fields can be generated within the oxide. Nevertheless, the ion concentration has an upper theoretical limit at the complete ionization of HfO\textsubscript{2}, while ion mobility cannot be too high so to avoid changing the distribution of ions when the read signal is applied. The MW also appears to have a weaker dependence on the concentration as compared to the mobility of the ions. Similarly, Figure \ref{fig:trade-off}b shows t\textsubscript{ret} as function of \textmu$_{\rm i}$ and c$_{\rm i}$ for t\textsubscript{w} = 1 \textmu s. Here, t\textsubscript{ret} decreases with increasing \textmu$_{\rm i}$ and c$_{\rm i}$ values, with quite similar behavior with respect to both quantities. Figures \ref{fig:trade-off}a and \ref{fig:trade-off}b also suggest an interdependence between \textmu$_{\rm i}$ and c$_{\rm i}$ to achieve the desired value of MW and t\textsubscript{ret}. Figures \ref{fig:tw}c-d and Fig. \ref{fig:trade-off}a-b clearly depict a trade-off between the two figures-of-merit of the ferroelectric-like FETs, as illustrated more clearly in Figs. \ref{fig:trade-off}c and \ref{fig:trade-off}d with respect to c$_{\rm i}$ and \textmu$_{\rm i}$, respectively. Therefore, a careful design is needed for a specific target application. 

\begin{figure}[!t]
	\begin{center}
		\includegraphics[width=1\linewidth]{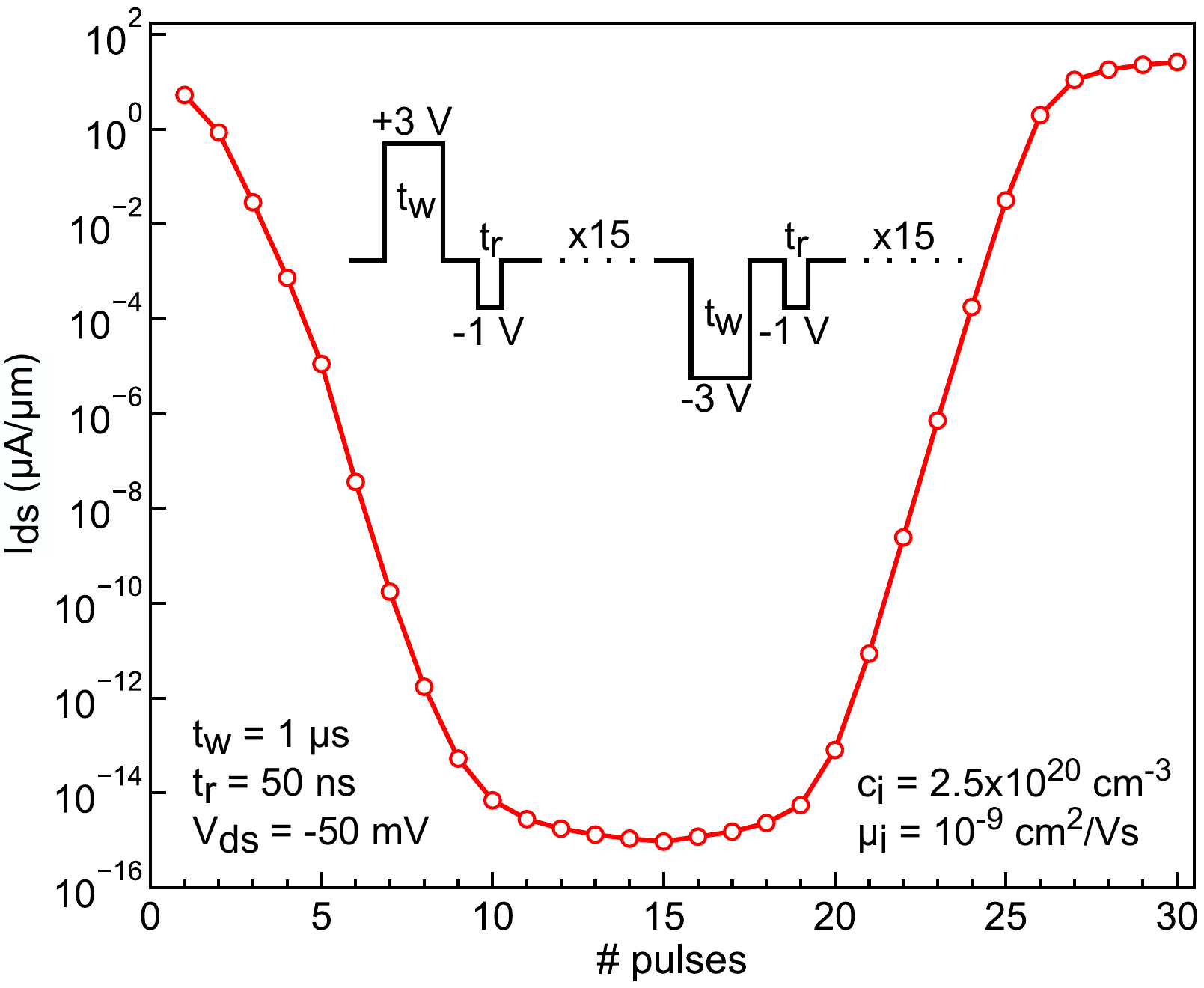}
		\caption{\label{fig:PPF_PPD} Response of the ferroelectric-like FET for a train of SET pulses followed by a train of RESET pulses. The series of SET pulses results into a gradual decrease of the current emulating paired-pulse depression (PPD) followed by a series of RESET pulses emulating the paired-pulse facilitation (PPF) functionality of the biological synapses.}
	\end{center}
\end{figure}

In addition to the V\textsubscript{T} change achieved by varying t\textsubscript{w} in Figure \ref{fig:tw}a, another way to continuously tune the V\textsubscript{T} is by repeatedly applying SET/RESET pulses of same amplitude and t\textsubscript{w} to emulate the paired-pulse facilitation (PPF) and paired-pulse depression (PPD) functions of the biological synapse. Figure \ref{fig:PPF_PPD} shows the variation of I\textsubscript{ds} at read voltages V\textsubscript{gs} = -1 V and V\textsubscript{ds} = -50 mV, after each consecutively applied SET/RESET pulse of t\textsubscript{w} = 1 \textmu s. The progressive application of SET pulses results into a PPD, where a change in the I\textsubscript{ds} of about 14 orders of magnitude could be achieved. The current can then be gradually increased (PPF) by the subsequent application of a similar amount of 15 consecutive RESET pulses, returning to a high level of current. Thus, the application of the appropriate set of writing and reading pulses, with suitable amplitudes and durations, induces a synaptic-like behavior on ferreoelectric-like FETs, an essential condition for their further implementation in ANNs.

\section{Conclusions}
This work presented an \textit{in-house} numerical tool able to self-consistently simulate the electrostatic and time-dependent electronic and ionic transports in iontronic devices. The tool is proved to successfully reproduce the experimental data, and is used to gain further insights into the operation of ferroelectric-like FETs. The dependence of the figures-of-merit, memory window and retention time, on the device physical parameters and their origin were analyzed. A trade-off was noticed between memory window and retention time with respect to the duration of the writing pulse, and the mobility and concentration of the ions. The change in the conductivity of the ferroelectric-like FETs in response to a train of pulses is also proved to emulate the paired-pulse facilitation and paired-pulse depression functionality of the biological synapses. In conclusion, voltage-driven ion migration not only tunes the conductance and threshold voltage of these devices, but also opens the possibility to create energy-efficient artificial neurons out of ferroelectric-like FETs based on amorphous gate oxides. In view of such promising results, their application on ANNs should be further pursued, and their operation optimised to foster the development of the neuromorphic computing paradigm.

\begin{acknowledgments}
	This work is supported by the Spanish Government through projects PID2020-116518GB-I00 funded by MCIN/AEI/10.13039/501100011033 and TED2021-129769B-I00 funded by MCIN/AEI/10.13039/501100011033 and the European Union NextGenerationEU/PRTR. This work is also supported by the R+D+i project A-ING-253-UGR23 AMBITIONS co-financed by Consejería de Universidad, Investigación e Innovación and the European Union under the FEDER Andalucía 2021-2027. J. Cuesta-Lopez acknowledges the FPU19/05132 program. M. D. Ganeriwala acknowledges funding from the European Union's Horizon 2020 research and innovation programme under the Marie Sklodowska-Curie grant agreement No. 101032701.
\end{acknowledgments}

\section*{Data Availability Statement}

The data that support the findings of this study are available from the corresponding author upon reasonable request.

\section*{Conflict of Interest Statement}

The authors have no conflicts to disclose.

\section*{Ethics Approval Statement}

Not applicable.

\section*{Author Contributions}

J.C.-L. and A.G. are the corresponding authors. J.C.-L. and A.T.-L. developed the simulator. J.C.-L., E.G.M., F.G.R. and F.P. conceived the numerical experiments. J.C.-L. and M.D.G. conducted the simulations and data processing. E.G.M., F.G.R. and A.G. acquired funding. J.C.-L., E.G.M. and M.D.G. wrote the original draft. All authors contributed to the revision and editing of the final manuscript. All authors have read and agreed to the published version of the manuscript.

\appendix*
\section{Details about the experimental validation}
This appendix summarizes the details about the calculation of the density of states, g(E), of Germanium, as well as the evaluation of interface traps.

\begin{figure}[t!]
\begin{center}
    \includegraphics[width=1\linewidth]{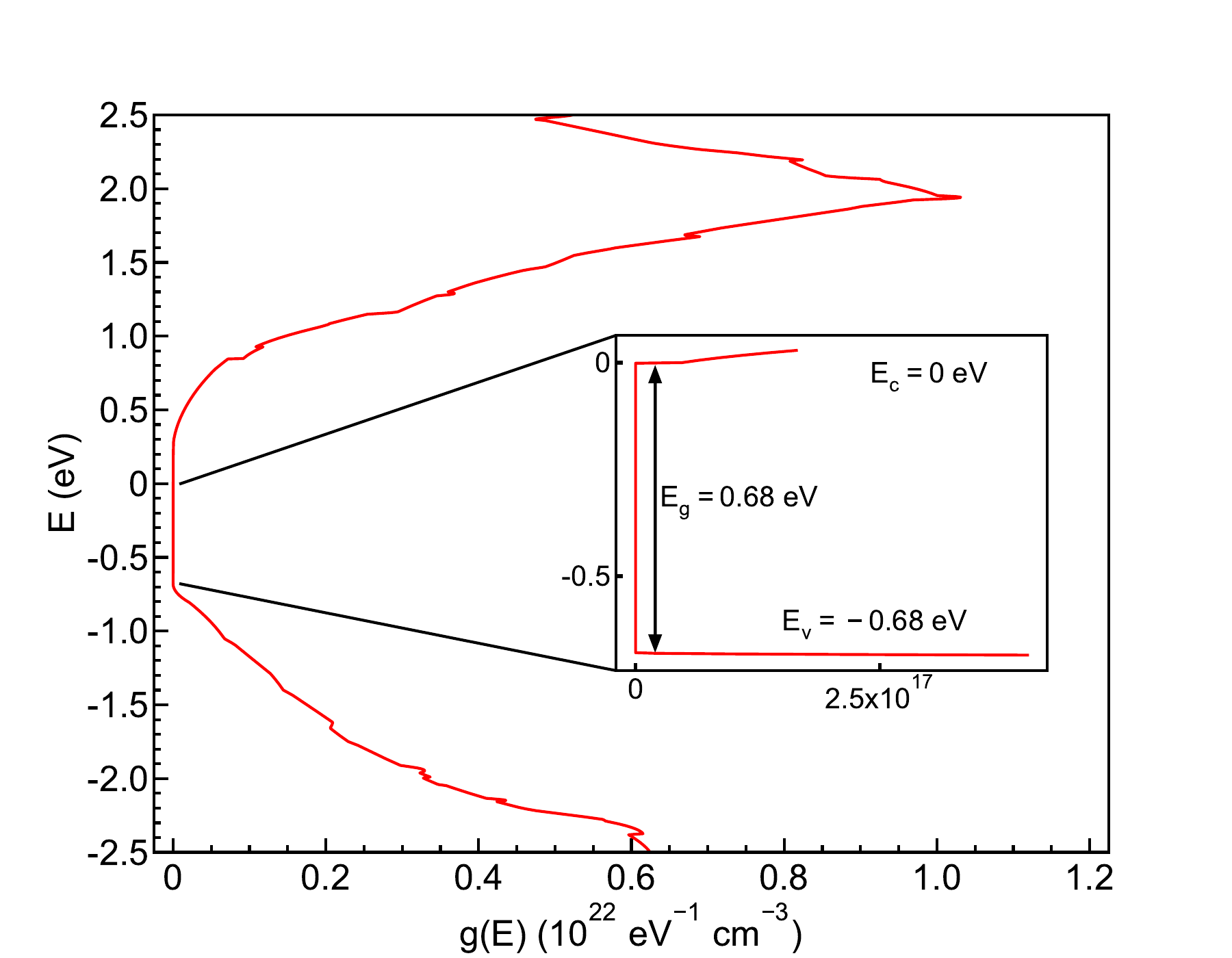}
    \caption{\label{fig:Ge} Germanium density of states, g(E), computed using the density functional theory (DFT) calculations. In particular, we used the Hybrid Generalized Gradient Approximation (Hybrid GGA) as implemented in QuantumATK with the LCAO basis and norm-conserving SG15 pseudo-potential. Brillouin-zone integration was performed over $9 \times 9 \times 9$ Monkhorst-Pack Grid. A sufficiently large energy cutt-off of 100 Hartree was used.}
\end{center}
\end{figure}

We have used the software QuantumATK to calculate the density of states of Germanium, g(E), by means of the Density Functional Theory (DFT), as represented in Figure \ref{fig:Ge}. The same g(E) is used in Equation \ref{eq:eh} to evaluate the concentration of free charge carriers, $n$ and $p$.

\begin{figure}[h!]
	\begin{center}
		\includegraphics[width=1\linewidth]{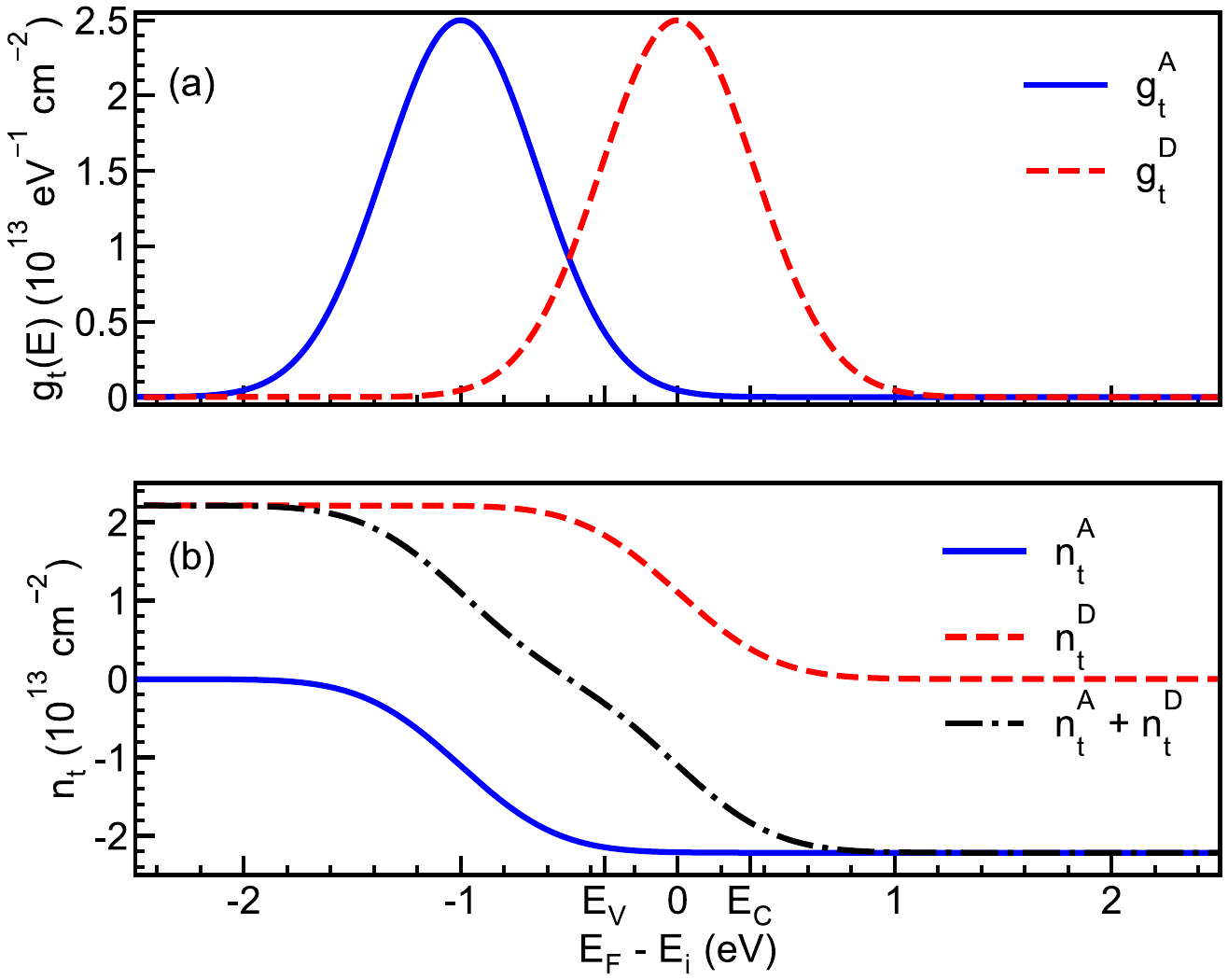}
		\caption{\label{fig:traps} (a) Trap density of states and (b) charged-trap density as a function of energy considering E$_{\rm i}=0$ eV. The concentration of each trap species follows a Gaussian density of states, g$_{\rm t,A} = 2.5 \cdot 10^{13} e^{- (\text{E} + 1)^2 / 0.5^2 }$ cm$^{-2}$ eV$^{-1}$ and g$_{\rm t,D} = 2.5 \cdot 10^{13} e^{- \text{E}^2 / 0.5^2 }$ cm$^{-2}$ eV$^{-1}$. The charged-trap density at a certain energy level is evaluated by integration over all of the trap energy states below. Acceptor traps, n$_{\rm t}^{\rm A}$, are neutral when unoccupied, and become negatively charged when occupied by the capture of an electron. Donor traps, n$_{\rm t}^{\rm D}$, are neutral when occupied, and become positively charged when unoccupied by the emission of an electron.
		}
	\end{center}
\end{figure}

Interface traps were also considered for the experimental validation carried out in Section \ref{sec:exp}. Traps were modeled in a narrow layer of barely 2 nm at the HfO$_2$/Ge interface. Their concentration, n$_{\rm t}$, was evaluated similarly to the electron and hole densities, i.e., by defining a Gaussian density of states around the trap energy levels that account for the energy values electrons should have to be captured or emitted by acceptor and donor traps, respectively. Further details are described in Figure \ref{fig:traps}. Finally, note that the energy reference was redefined to be E$_{\rm i}$, the intrinsic Fermi energy level, instead of E$_{\rm c}$, the bottom of the conduction band.

\bibliography{manuscript_preprint}

\begin{thebibliography}{45}%
\makeatletter
\providecommand \@ifxundefined [1]{%
 \@ifx{#1\undefined}
}%
\providecommand \@ifnum [1]{%
 \ifnum #1\expandafter \@firstoftwo
 \else \expandafter \@secondoftwo
 \fi
}%
\providecommand \@ifx [1]{%
 \ifx #1\expandafter \@firstoftwo
 \else \expandafter \@secondoftwo
 \fi
}%
\providecommand \natexlab [1]{#1}%
\providecommand \enquote  [1]{``#1''}%
\providecommand \bibnamefont  [1]{#1}%
\providecommand \bibfnamefont [1]{#1}%
\providecommand \citenamefont [1]{#1}%
\providecommand \href@noop [0]{\@secondoftwo}%
\providecommand \href [0]{\begingroup \@sanitize@url \@href}%
\providecommand \@href[1]{\@@startlink{#1}\@@href}%
\providecommand \@@href[1]{\endgroup#1\@@endlink}%
\providecommand \@sanitize@url [0]{\catcode `\\12\catcode `\$12\catcode
  `\&12\catcode `\#12\catcode `\^12\catcode `\_12\catcode `\%12\relax}%
\providecommand \@@startlink[1]{}%
\providecommand \@@endlink[0]{}%
\providecommand \url  [0]{\begingroup\@sanitize@url \@url }%
\providecommand \@url [1]{\endgroup\@href {#1}{\urlprefix }}%
\providecommand \urlprefix  [0]{URL }%
\providecommand \Eprint [0]{\href }%
\providecommand \doibase [0]{http://dx.doi.org/}%
\providecommand \selectlanguage [0]{\@gobble}%
\providecommand \bibinfo  [0]{\@secondoftwo}%
\providecommand \bibfield  [0]{\@secondoftwo}%
\providecommand \translation [1]{[#1]}%
\providecommand \BibitemOpen [0]{}%
\providecommand \bibitemStop [0]{}%
\providecommand \bibitemNoStop [0]{.\EOS\space}%
\providecommand \EOS [0]{\spacefactor3000\relax}%
\providecommand \BibitemShut  [1]{\csname bibitem#1\endcsname}%
\let\auto@bib@innerbib\@empty
\bibitem [{\citenamefont {Roy}, \citenamefont {Jaiswal},\ and\ \citenamefont
  {Panda}(2019)}]{Roy2019}%
  \BibitemOpen
  \bibfield  {author} {\bibinfo {author} {\bibfnamefont {K.}~\bibnamefont
  {Roy}}, \bibinfo {author} {\bibfnamefont {A.}~\bibnamefont {Jaiswal}}, \ and\
  \bibinfo {author} {\bibfnamefont {P.}~\bibnamefont {Panda}},\ }\bibfield
  {title} {\enquote {\bibinfo {title} {Towards spike-based machine intelligence
  with neuromorphic computing},}\ }\href@noop {} {\bibfield  {journal}
  {\bibinfo  {journal} {Nature}\ }\textbf {\bibinfo {volume} {575}},\ \bibinfo
  {pages} {607--617} (\bibinfo {year} {2019})}\BibitemShut {NoStop}%
\bibitem [{\citenamefont {Najafabadi}\ \emph {et~al.}(2015)\citenamefont
  {Najafabadi}, \citenamefont {Villanustre}, \citenamefont {Khoshgoftaar},
  \citenamefont {Seliya}, \citenamefont {Wald},\ and\ \citenamefont
  {Muharemagic}}]{Najafabadi2015}%
  \BibitemOpen
  \bibfield  {author} {\bibinfo {author} {\bibfnamefont {M.~M.}\ \bibnamefont
  {Najafabadi}}, \bibinfo {author} {\bibfnamefont {F.}~\bibnamefont
  {Villanustre}}, \bibinfo {author} {\bibfnamefont {T.~M.}\ \bibnamefont
  {Khoshgoftaar}}, \bibinfo {author} {\bibfnamefont {N.}~\bibnamefont
  {Seliya}}, \bibinfo {author} {\bibfnamefont {R.}~\bibnamefont {Wald}}, \ and\
  \bibinfo {author} {\bibfnamefont {E.}~\bibnamefont {Muharemagic}},\
  }\bibfield  {title} {\enquote {\bibinfo {title} {{Deep learning applications
  and challenges in big data analytics}},}\ }\href {\doibase
  10.1186/s40537-014-0007-7} {\bibfield  {journal} {\bibinfo  {journal}
  {Journal of Big Data}\ }\textbf {\bibinfo {volume} {2}},\ \bibinfo {pages}
  {1--21} (\bibinfo {year} {2015})}\BibitemShut {NoStop}%
\bibitem [{\citenamefont {Mead}\ and\ \citenamefont {Ismail}(1989)}]{Mead1989}%
  \BibitemOpen
  \bibfield  {author} {\bibinfo {author} {\bibfnamefont {C.}~\bibnamefont
  {Mead}}\ and\ \bibinfo {author} {\bibfnamefont {M.}~\bibnamefont {Ismail}},\
  }\href@noop {} {\emph {\bibinfo {title} {Analog VLSI implementation of neural
  systems}}},\ Vol.~\bibinfo {volume} {80}\ (\bibinfo  {publisher} {Springer
  Science \& Business Media},\ \bibinfo {year} {1989})\BibitemShut {NoStop}%
\bibitem [{\citenamefont {Zhu}\ \emph {et~al.}(2020)\citenamefont {Zhu},
  \citenamefont {Zhang}, \citenamefont {Yang},\ and\ \citenamefont
  {Huang}}]{Zhu2020}%
  \BibitemOpen
  \bibfield  {author} {\bibinfo {author} {\bibfnamefont {J.}~\bibnamefont
  {Zhu}}, \bibinfo {author} {\bibfnamefont {T.}~\bibnamefont {Zhang}}, \bibinfo
  {author} {\bibfnamefont {Y.}~\bibnamefont {Yang}}, \ and\ \bibinfo {author}
  {\bibfnamefont {R.}~\bibnamefont {Huang}},\ }\bibfield  {title} {\enquote
  {\bibinfo {title} {A comprehensive review on emerging artificial neuromorphic
  devices},}\ }\href@noop {} {\bibfield  {journal} {\bibinfo  {journal}
  {Applied Physics Reviews}\ }\textbf {\bibinfo {volume} {7}} (\bibinfo {year}
  {2020})}\BibitemShut {NoStop}%
\bibitem [{\citenamefont {Hebb}(1949)}]{Hebb1949}%
  \BibitemOpen
  \bibfield  {author} {\bibinfo {author} {\bibfnamefont {D.}~\bibnamefont
  {Hebb}},\ }\bibfield  {title} {\enquote {\bibinfo {title} {The organization
  of behavior; a neuropsychological theory.}}\ }\href@noop {} {\  (\bibinfo
  {year} {1949})}\BibitemShut {NoStop}%
\bibitem [{\citenamefont {Indiveri}\ and\ \citenamefont
  {Liu}(2015)}]{Indiveri2015}%
  \BibitemOpen
  \bibfield  {author} {\bibinfo {author} {\bibfnamefont {G.}~\bibnamefont
  {Indiveri}}\ and\ \bibinfo {author} {\bibfnamefont {S.~C.}\ \bibnamefont
  {Liu}},\ }\bibfield  {title} {\enquote {\bibinfo {title} {{Memory and
  Information Processing in Neuromorphic Systems}},}\ }\href {\doibase
  10.1109/JPROC.2015.2444094} {\bibfield  {journal} {\bibinfo  {journal}
  {Proceedings of the IEEE}\ }\textbf {\bibinfo {volume} {103}},\ \bibinfo
  {pages} {1379--1397} (\bibinfo {year} {2015})}\BibitemShut {NoStop}%
\bibitem [{\citenamefont {Zidan}, \citenamefont {Strachan},\ and\ \citenamefont
  {Lu}(2018)}]{Zidan2018}%
  \BibitemOpen
  \bibfield  {author} {\bibinfo {author} {\bibfnamefont {M.~A.}\ \bibnamefont
  {Zidan}}, \bibinfo {author} {\bibfnamefont {J.~P.}\ \bibnamefont {Strachan}},
  \ and\ \bibinfo {author} {\bibfnamefont {W.~D.}\ \bibnamefont {Lu}},\
  }\bibfield  {title} {\enquote {\bibinfo {title} {{The future of electronics
  based on memristive systems}},}\ }\href {\doibase 10.1038/s41928-017-0006-8}
  {\bibfield  {journal} {\bibinfo  {journal} {Nature Electronics}\ }\textbf
  {\bibinfo {volume} {1}},\ \bibinfo {pages} {22--29} (\bibinfo {year}
  {2018})}\BibitemShut {NoStop}%
\bibitem [{\citenamefont {Ielmini}(2011)}]{Ielmini2011}%
  \BibitemOpen
  \bibfield  {author} {\bibinfo {author} {\bibfnamefont {D.}~\bibnamefont
  {Ielmini}},\ }\bibfield  {title} {\enquote {\bibinfo {title} {{Modeling the
  Universal Set / Reset Characteristics of Filament Growth}},}\ }\href@noop {}
  {\bibfield  {journal} {\bibinfo  {journal} {IEEE Transactions on Electron
  Devices}\ }\textbf {\bibinfo {volume} {58}},\ \bibinfo {pages} {1--9}
  (\bibinfo {year} {2011})}\BibitemShut {NoStop}%
\bibitem [{\citenamefont {Larentis}\ \emph {et~al.}(2012)\citenamefont
  {Larentis}, \citenamefont {Nardi}, \citenamefont {Balatti}, \citenamefont
  {Gilmer},\ and\ \citenamefont {Ielmini}}]{Larentis2012}%
  \BibitemOpen
  \bibfield  {author} {\bibinfo {author} {\bibfnamefont {S.}~\bibnamefont
  {Larentis}}, \bibinfo {author} {\bibfnamefont {F.}~\bibnamefont {Nardi}},
  \bibinfo {author} {\bibfnamefont {S.}~\bibnamefont {Balatti}}, \bibinfo
  {author} {\bibfnamefont {D.~C.}\ \bibnamefont {Gilmer}}, \ and\ \bibinfo
  {author} {\bibfnamefont {D.}~\bibnamefont {Ielmini}},\ }\bibfield  {title}
  {\enquote {\bibinfo {title} {{Resistive switching by voltage-driven ion
  migration in bipolar RRAMPart II: Modeling}},}\ }\href {\doibase
  10.1109/TED.2012.2202320} {\bibfield  {journal} {\bibinfo  {journal} {IEEE
  Transactions on Electron Devices}\ }\textbf {\bibinfo {volume} {59}},\
  \bibinfo {pages} {2468--2475} (\bibinfo {year} {2012})}\BibitemShut {NoStop}%
\bibitem [{\citenamefont {Wang}\ \emph {et~al.}(2023)\citenamefont {Wang},
  \citenamefont {Shi}, \citenamefont {Qiao}, \citenamefont {Lin}, \citenamefont
  {Wu},\ and\ \citenamefont {Hu}}]{Wang2023}%
  \BibitemOpen
  \bibfield  {author} {\bibinfo {author} {\bibfnamefont {C.}~\bibnamefont
  {Wang}}, \bibinfo {author} {\bibfnamefont {G.}~\bibnamefont {Shi}}, \bibinfo
  {author} {\bibfnamefont {F.}~\bibnamefont {Qiao}}, \bibinfo {author}
  {\bibfnamefont {R.}~\bibnamefont {Lin}}, \bibinfo {author} {\bibfnamefont
  {S.}~\bibnamefont {Wu}}, \ and\ \bibinfo {author} {\bibfnamefont
  {Z.}~\bibnamefont {Hu}},\ }\bibfield  {title} {\enquote {\bibinfo {title}
  {{Research progress in architecture and application of RRAM with
  computing-in-memory}},}\ }\href {\doibase 10.1039/d3na00025g} {\bibfield
  {journal} {\bibinfo  {journal} {Nanoscale Advances}\ ,\ \bibinfo {pages}
  {1559--1573}} (\bibinfo {year} {2023})}\BibitemShut {NoStop}%
\bibitem [{\citenamefont {Zhu}\ \emph {et~al.}(2018)\citenamefont {Zhu},
  \citenamefont {Yang}, \citenamefont {Jia}, \citenamefont {Liang},
  \citenamefont {Zhu}, \citenamefont {Rehman}, \citenamefont {Bao},
  \citenamefont {Zhang}, \citenamefont {Cai}, \citenamefont {Song},\ and\
  \citenamefont {Huang}}]{Zhu2018}%
  \BibitemOpen
  \bibfield  {author} {\bibinfo {author} {\bibfnamefont {J.}~\bibnamefont
  {Zhu}}, \bibinfo {author} {\bibfnamefont {Y.}~\bibnamefont {Yang}}, \bibinfo
  {author} {\bibfnamefont {R.}~\bibnamefont {Jia}}, \bibinfo {author}
  {\bibfnamefont {Z.}~\bibnamefont {Liang}}, \bibinfo {author} {\bibfnamefont
  {W.}~\bibnamefont {Zhu}}, \bibinfo {author} {\bibfnamefont {Z.~U.}\
  \bibnamefont {Rehman}}, \bibinfo {author} {\bibfnamefont {L.}~\bibnamefont
  {Bao}}, \bibinfo {author} {\bibfnamefont {X.}~\bibnamefont {Zhang}}, \bibinfo
  {author} {\bibfnamefont {Y.}~\bibnamefont {Cai}}, \bibinfo {author}
  {\bibfnamefont {L.}~\bibnamefont {Song}}, \ and\ \bibinfo {author}
  {\bibfnamefont {R.}~\bibnamefont {Huang}},\ }\bibfield  {title} {\enquote
  {\bibinfo {title} {{Ion Gated Synaptic Transistors Based on 2D van der Waals
  Crystals with Tunable Diffusive Dynamics}},}\ }\href {\doibase
  10.1002/adma.201800195} {\bibfield  {journal} {\bibinfo  {journal} {Advanced
  Materials}\ }\textbf {\bibinfo {volume} {30}},\ \bibinfo {pages} {1--11}
  (\bibinfo {year} {2018})}\BibitemShut {NoStop}%
\bibitem [{\citenamefont {Fuller}\ \emph {et~al.}(2019)\citenamefont {Fuller},
  \citenamefont {Li}, \citenamefont {Bennet}, \citenamefont {Keene},
  \citenamefont {Melianas}, \citenamefont {Agarwal}, \citenamefont {Marinella},
  \citenamefont {Salleo},\ and\ \citenamefont {Talin}}]{Fuller2019}%
  \BibitemOpen
  \bibfield  {author} {\bibinfo {author} {\bibfnamefont {E.~J.}\ \bibnamefont
  {Fuller}}, \bibinfo {author} {\bibfnamefont {Y.}~\bibnamefont {Li}}, \bibinfo
  {author} {\bibfnamefont {C.}~\bibnamefont {Bennet}}, \bibinfo {author}
  {\bibfnamefont {S.~T.}\ \bibnamefont {Keene}}, \bibinfo {author}
  {\bibfnamefont {A.}~\bibnamefont {Melianas}}, \bibinfo {author}
  {\bibfnamefont {S.}~\bibnamefont {Agarwal}}, \bibinfo {author} {\bibfnamefont
  {M.~J.}\ \bibnamefont {Marinella}}, \bibinfo {author} {\bibfnamefont
  {A.}~\bibnamefont {Salleo}}, \ and\ \bibinfo {author} {\bibfnamefont {A.~A.}\
  \bibnamefont {Talin}},\ }\bibfield  {title} {\enquote {\bibinfo {title}
  {Redox transistors for neuromorphic computing},}\ }\href {\doibase
  10.1147/JRD.2019.2942285} {\bibfield  {journal} {\bibinfo  {journal} {IBM
  Journal of Research and Development}\ }\textbf {\bibinfo {volume} {63}},\
  \bibinfo {pages} {9:1--9:9} (\bibinfo {year} {2019})}\BibitemShut {NoStop}%
\bibitem [{\citenamefont {Jerry}\ \emph {et~al.}(2017)\citenamefont {Jerry},
  \citenamefont {Chen}, \citenamefont {Zhang}, \citenamefont {Sharma},
  \citenamefont {Ni}, \citenamefont {Yu},\ and\ \citenamefont
  {Datta}}]{Jerry2017}%
  \BibitemOpen
  \bibfield  {author} {\bibinfo {author} {\bibfnamefont {M.}~\bibnamefont
  {Jerry}}, \bibinfo {author} {\bibfnamefont {P.-Y.}\ \bibnamefont {Chen}},
  \bibinfo {author} {\bibfnamefont {J.}~\bibnamefont {Zhang}}, \bibinfo
  {author} {\bibfnamefont {P.}~\bibnamefont {Sharma}}, \bibinfo {author}
  {\bibfnamefont {K.}~\bibnamefont {Ni}}, \bibinfo {author} {\bibfnamefont
  {S.}~\bibnamefont {Yu}}, \ and\ \bibinfo {author} {\bibfnamefont
  {S.}~\bibnamefont {Datta}},\ }\bibfield  {title} {\enquote {\bibinfo {title}
  {Ferroelectric fet analog synapse for acceleration of deep neural network
  training},}\ }in\ \href {\doibase 10.1109/IEDM.2017.8268338} {\emph {\bibinfo
  {booktitle} {2017 IEEE International Electron Devices Meeting (IEDM)}}}\
  (\bibinfo {year} {2017})\ pp.\ \bibinfo {pages} {6.2.1--6.2.4}\BibitemShut
  {NoStop}%
\bibitem [{\citenamefont {Si}\ \emph {et~al.}(2019)\citenamefont {Si},
  \citenamefont {Saha}, \citenamefont {Gao}, \citenamefont {Qiu}, \citenamefont
  {Qin}, \citenamefont {Duan}, \citenamefont {Jian}, \citenamefont {Niu},
  \citenamefont {Wang}, \citenamefont {Wu}, \citenamefont {Gupta},\ and\
  \citenamefont {Ye}}]{Si2019}%
  \BibitemOpen
  \bibfield  {author} {\bibinfo {author} {\bibfnamefont {M.}~\bibnamefont
  {Si}}, \bibinfo {author} {\bibfnamefont {A.~K.}\ \bibnamefont {Saha}},
  \bibinfo {author} {\bibfnamefont {S.}~\bibnamefont {Gao}}, \bibinfo {author}
  {\bibfnamefont {G.}~\bibnamefont {Qiu}}, \bibinfo {author} {\bibfnamefont
  {J.}~\bibnamefont {Qin}}, \bibinfo {author} {\bibfnamefont {Y.}~\bibnamefont
  {Duan}}, \bibinfo {author} {\bibfnamefont {J.}~\bibnamefont {Jian}}, \bibinfo
  {author} {\bibfnamefont {C.}~\bibnamefont {Niu}}, \bibinfo {author}
  {\bibfnamefont {H.}~\bibnamefont {Wang}}, \bibinfo {author} {\bibfnamefont
  {W.}~\bibnamefont {Wu}}, \bibinfo {author} {\bibfnamefont {S.~K.}\
  \bibnamefont {Gupta}}, \ and\ \bibinfo {author} {\bibfnamefont {P.~D.}\
  \bibnamefont {Ye}},\ }\bibfield  {title} {\enquote {\bibinfo {title} {{A
  ferroelectric semiconductor field-effect transistor}},}\ }\href {\doibase
  10.1038/s41928-019-0338-7} {\bibfield  {journal} {\bibinfo  {journal} {Nature
  Electronics}\ }\textbf {\bibinfo {volume} {2}},\ \bibinfo {pages} {580--586}
  (\bibinfo {year} {2019})}\BibitemShut {NoStop}%
\bibitem [{\citenamefont {Chen}\ \emph {et~al.}(2020)\citenamefont {Chen},
  \citenamefont {Wang}, \citenamefont {Peng}, \citenamefont {Feng},
  \citenamefont {Sarkar}, \citenamefont {Li}, \citenamefont {Li}, \citenamefont
  {Liu}, \citenamefont {Han}, \citenamefont {Gong}, \citenamefont {Chen},
  \citenamefont {Liu}, \citenamefont {Han},\ and\ \citenamefont
  {Ang}}]{Chen2020}%
  \BibitemOpen
  \bibfield  {author} {\bibinfo {author} {\bibfnamefont {L.}~\bibnamefont
  {Chen}}, \bibinfo {author} {\bibfnamefont {L.}~\bibnamefont {Wang}}, \bibinfo
  {author} {\bibfnamefont {Y.}~\bibnamefont {Peng}}, \bibinfo {author}
  {\bibfnamefont {X.}~\bibnamefont {Feng}}, \bibinfo {author} {\bibfnamefont
  {S.}~\bibnamefont {Sarkar}}, \bibinfo {author} {\bibfnamefont
  {S.}~\bibnamefont {Li}}, \bibinfo {author} {\bibfnamefont {B.}~\bibnamefont
  {Li}}, \bibinfo {author} {\bibfnamefont {L.}~\bibnamefont {Liu}}, \bibinfo
  {author} {\bibfnamefont {K.}~\bibnamefont {Han}}, \bibinfo {author}
  {\bibfnamefont {X.}~\bibnamefont {Gong}}, \bibinfo {author} {\bibfnamefont
  {J.}~\bibnamefont {Chen}}, \bibinfo {author} {\bibfnamefont {Y.}~\bibnamefont
  {Liu}}, \bibinfo {author} {\bibfnamefont {G.}~\bibnamefont {Han}}, \ and\
  \bibinfo {author} {\bibfnamefont {K.~W.}\ \bibnamefont {Ang}},\ }\bibfield
  {title} {\enquote {\bibinfo {title} {{A van der Waals Synaptic Transistor
  Based on Ferroelectric Hf$_{0.5}$Zr$_{0.5}$O2 and 2D Tungsten Disulfide}},}\
  }\href {\doibase 10.1002/aelm.202000057} {\bibfield  {journal} {\bibinfo
  {journal} {Advanced Electronic Materials}\ }\textbf {\bibinfo {volume} {6}},\
  \bibinfo {pages} {1--9} (\bibinfo {year} {2020})}\BibitemShut {NoStop}%
\bibitem [{\citenamefont {Kim}, \citenamefont {Choi},\ and\ \citenamefont
  {Jang}(2021)}]{Kim2021}%
  \BibitemOpen
  \bibfield  {author} {\bibinfo {author} {\bibfnamefont {J.~Y.}\ \bibnamefont
  {Kim}}, \bibinfo {author} {\bibfnamefont {M.~J.}\ \bibnamefont {Choi}}, \
  and\ \bibinfo {author} {\bibfnamefont {H.~W.}\ \bibnamefont {Jang}},\
  }\bibfield  {title} {\enquote {\bibinfo {title} {{Ferroelectric field effect
  transistors: Progress and perspective}},}\ }\href {\doibase
  10.1063/5.0035515} {\bibfield  {journal} {\bibinfo  {journal} {APL
  Materials}\ }\textbf {\bibinfo {volume} {9}},\ \bibinfo {pages} {1--18}
  (\bibinfo {year} {2021})}\BibitemShut {NoStop}%
\bibitem [{\citenamefont {Chatterjee}\ \emph {et~al.}(2017)\citenamefont
  {Chatterjee}, \citenamefont {Kim}, \citenamefont {Karbasian}, \citenamefont
  {Tan}, \citenamefont {Yadav}, \citenamefont {Khan}, \citenamefont {Hu},\ and\
  \citenamefont {Salahuddin}}]{Chatterjee2017}%
  \BibitemOpen
  \bibfield  {author} {\bibinfo {author} {\bibfnamefont {K.}~\bibnamefont
  {Chatterjee}}, \bibinfo {author} {\bibfnamefont {S.}~\bibnamefont {Kim}},
  \bibinfo {author} {\bibfnamefont {G.}~\bibnamefont {Karbasian}}, \bibinfo
  {author} {\bibfnamefont {A.~J.}\ \bibnamefont {Tan}}, \bibinfo {author}
  {\bibfnamefont {A.~K.}\ \bibnamefont {Yadav}}, \bibinfo {author}
  {\bibfnamefont {A.~I.}\ \bibnamefont {Khan}}, \bibinfo {author}
  {\bibfnamefont {C.}~\bibnamefont {Hu}}, \ and\ \bibinfo {author}
  {\bibfnamefont {S.}~\bibnamefont {Salahuddin}},\ }\bibfield  {title}
  {\enquote {\bibinfo {title} {{Self-Aligned, Gate Last, FDSOI, Ferroelectric
  Gate Memory Device With 5.5-nm HF$_{0.8}$Zr$_{0.2}$O$_2$, High Endurance and
  Breakdown Recovery}},}\ }\href {\doibase 10.1109/LED.2017.2748992} {\bibfield
   {journal} {\bibinfo  {journal} {IEEE Electron Device Letters}\ }\textbf
  {\bibinfo {volume} {38}},\ \bibinfo {pages} {1379--1382} (\bibinfo {year}
  {2017})}\BibitemShut {NoStop}%
\bibitem [{\citenamefont {Ni}\ \emph {et~al.}(2018)\citenamefont {Ni},
  \citenamefont {Sharma}, \citenamefont {Zhang}, \citenamefont {Jerry},
  \citenamefont {Smith}, \citenamefont {Tapily}, \citenamefont {Clark},
  \citenamefont {Mahapatra},\ and\ \citenamefont {Datta}}]{Ni2018}%
  \BibitemOpen
  \bibfield  {author} {\bibinfo {author} {\bibfnamefont {K.}~\bibnamefont
  {Ni}}, \bibinfo {author} {\bibfnamefont {P.}~\bibnamefont {Sharma}}, \bibinfo
  {author} {\bibfnamefont {J.}~\bibnamefont {Zhang}}, \bibinfo {author}
  {\bibfnamefont {M.}~\bibnamefont {Jerry}}, \bibinfo {author} {\bibfnamefont
  {J.~A.}\ \bibnamefont {Smith}}, \bibinfo {author} {\bibfnamefont
  {K.}~\bibnamefont {Tapily}}, \bibinfo {author} {\bibfnamefont
  {R.}~\bibnamefont {Clark}}, \bibinfo {author} {\bibfnamefont
  {S.}~\bibnamefont {Mahapatra}}, \ and\ \bibinfo {author} {\bibfnamefont
  {S.}~\bibnamefont {Datta}},\ }\bibfield  {title} {\enquote {\bibinfo {title}
  {{Critical Role of Interlayer in Hf$_{0.5}$Zr$_{0.5}$O$_2$ Ferroelectric FET
  Nonvolatile Memory Performance}},}\ }\href {\doibase
  10.1109/TED.2018.2829122} {\bibfield  {journal} {\bibinfo  {journal} {IEEE
  Transactions on Electron Devices}\ }\textbf {\bibinfo {volume} {65}},\
  \bibinfo {pages} {2461--2469} (\bibinfo {year} {2018})}\BibitemShut {NoStop}%
\bibitem [{\citenamefont {Osada}\ and\ \citenamefont {Sasaki}(2019)}]{Osada19}%
  \BibitemOpen
  \bibfield  {author} {\bibinfo {author} {\bibfnamefont {M.}~\bibnamefont
  {Osada}}\ and\ \bibinfo {author} {\bibfnamefont {T.}~\bibnamefont {Sasaki}},\
  }\bibfield  {title} {\enquote {\bibinfo {title} {{The rise of 2D
  dielectrics/ferroelectrics}},}\ }\href {\doibase 10.1063/1.5129447}
  {\bibfield  {journal} {\bibinfo  {journal} {APL Materials}\ }\textbf
  {\bibinfo {volume} {7}},\ \bibinfo {pages} {120902} (\bibinfo {year}
  {2019})}\BibitemShut {NoStop}%
\bibitem [{\citenamefont {Muller}\ \emph {et~al.}(2013)\citenamefont {Muller},
  \citenamefont {Boscke}, \citenamefont {Muller}, \citenamefont {Yurchuk},
  \citenamefont {Polakowski}, \citenamefont {Paul}, \citenamefont {Martin},
  \citenamefont {Schenk}, \citenamefont {Khullar}, \citenamefont {Kersch},
  \citenamefont {Weinreich}, \citenamefont {Riedel}, \citenamefont {Seidel},
  \citenamefont {Kumar}, \citenamefont {Arruda}, \citenamefont {Kalinin},
  \citenamefont {Schlosser}, \citenamefont {Boschke}, \citenamefont {{Van
  Bentum}}, \citenamefont {Schroder},\ and\ \citenamefont
  {Mikolajick}}]{Muller2013}%
  \BibitemOpen
  \bibfield  {author} {\bibinfo {author} {\bibfnamefont {J.}~\bibnamefont
  {Muller}}, \bibinfo {author} {\bibfnamefont {T.~S.}\ \bibnamefont {Boscke}},
  \bibinfo {author} {\bibfnamefont {S.}~\bibnamefont {Muller}}, \bibinfo
  {author} {\bibfnamefont {E.}~\bibnamefont {Yurchuk}}, \bibinfo {author}
  {\bibfnamefont {P.}~\bibnamefont {Polakowski}}, \bibinfo {author}
  {\bibfnamefont {J.}~\bibnamefont {Paul}}, \bibinfo {author} {\bibfnamefont
  {D.}~\bibnamefont {Martin}}, \bibinfo {author} {\bibfnamefont
  {T.}~\bibnamefont {Schenk}}, \bibinfo {author} {\bibfnamefont
  {K.}~\bibnamefont {Khullar}}, \bibinfo {author} {\bibfnamefont
  {A.}~\bibnamefont {Kersch}}, \bibinfo {author} {\bibfnamefont
  {W.}~\bibnamefont {Weinreich}}, \bibinfo {author} {\bibfnamefont
  {S.}~\bibnamefont {Riedel}}, \bibinfo {author} {\bibfnamefont
  {K.}~\bibnamefont {Seidel}}, \bibinfo {author} {\bibfnamefont
  {A.}~\bibnamefont {Kumar}}, \bibinfo {author} {\bibfnamefont {T.~M.}\
  \bibnamefont {Arruda}}, \bibinfo {author} {\bibfnamefont {S.~V.}\
  \bibnamefont {Kalinin}}, \bibinfo {author} {\bibfnamefont {T.}~\bibnamefont
  {Schlosser}}, \bibinfo {author} {\bibfnamefont {R.}~\bibnamefont {Boschke}},
  \bibinfo {author} {\bibfnamefont {R.}~\bibnamefont {{Van Bentum}}}, \bibinfo
  {author} {\bibfnamefont {U.}~\bibnamefont {Schroder}}, \ and\ \bibinfo
  {author} {\bibfnamefont {T.}~\bibnamefont {Mikolajick}},\ }\bibfield  {title}
  {\enquote {\bibinfo {title} {{Ferroelectric hafnium oxide: A CMOS-compatible
  and highly scalable approach to future ferroelectric memories}},}\ }\href
  {\doibase 10.1109/IEDM.2013.6724605} {\bibfield  {journal} {\bibinfo
  {journal} {Technical Digest - International Electron Devices Meeting, IEDM}\
  ,\ \bibinfo {pages} {280--283}} (\bibinfo {year} {2013})}\BibitemShut
  {NoStop}%
\bibitem [{\citenamefont {Liu}\ \emph {et~al.}(2019)\citenamefont {Liu},
  \citenamefont {Wang}, \citenamefont {Han}, \citenamefont {Li}, \citenamefont
  {Peng}, \citenamefont {Liu}, \citenamefont {Wang}, \citenamefont {Zhong},
  \citenamefont {Duan}, \citenamefont {Wang}, \citenamefont {Xu}, \citenamefont
  {Liu},\ and\ \citenamefont {Hao}}]{Liu2019}%
  \BibitemOpen
  \bibfield  {author} {\bibinfo {author} {\bibfnamefont {H.}~\bibnamefont
  {Liu}}, \bibinfo {author} {\bibfnamefont {C.}~\bibnamefont {Wang}}, \bibinfo
  {author} {\bibfnamefont {G.}~\bibnamefont {Han}}, \bibinfo {author}
  {\bibfnamefont {J.}~\bibnamefont {Li}}, \bibinfo {author} {\bibfnamefont
  {Y.}~\bibnamefont {Peng}}, \bibinfo {author} {\bibfnamefont {Y.}~\bibnamefont
  {Liu}}, \bibinfo {author} {\bibfnamefont {X.}~\bibnamefont {Wang}}, \bibinfo
  {author} {\bibfnamefont {N.}~\bibnamefont {Zhong}}, \bibinfo {author}
  {\bibfnamefont {C.}~\bibnamefont {Duan}}, \bibinfo {author} {\bibfnamefont
  {X.}~\bibnamefont {Wang}}, \bibinfo {author} {\bibfnamefont {N.}~\bibnamefont
  {Xu}}, \bibinfo {author} {\bibfnamefont {T.-J.~K.}\ \bibnamefont {Liu}}, \
  and\ \bibinfo {author} {\bibfnamefont {Y.}~\bibnamefont {Hao}},\ }\bibfield
  {title} {\enquote {\bibinfo {title} {{ZrO$_2$ Ferroelectric FET for
  Non-volatile Memory Application}},}\ }\href {\doibase
  10.1109/led.2019.2930458} {\bibfield  {journal} {\bibinfo  {journal} {IEEE
  Electron Device Letters}\ }\textbf {\bibinfo {volume} {40}},\ \bibinfo
  {pages} {1419--1422} (\bibinfo {year} {2019})}\BibitemShut {NoStop}%
\bibitem [{\citenamefont {Peng}\ \emph
  {et~al.}(2020{\natexlab{a}})\citenamefont {Peng}, \citenamefont {Xiao},
  \citenamefont {Han}, \citenamefont {Liu}, \citenamefont {Liu}, \citenamefont
  {Liu}, \citenamefont {Zhou}, \citenamefont {Yang}, \citenamefont {Zhong},
  \citenamefont {Duan},\ and\ \citenamefont {Hao}}]{Peng2020b}%
  \BibitemOpen
  \bibfield  {author} {\bibinfo {author} {\bibfnamefont {Y.}~\bibnamefont
  {Peng}}, \bibinfo {author} {\bibfnamefont {W.}~\bibnamefont {Xiao}}, \bibinfo
  {author} {\bibfnamefont {G.}~\bibnamefont {Han}}, \bibinfo {author}
  {\bibfnamefont {Y.}~\bibnamefont {Liu}}, \bibinfo {author} {\bibfnamefont
  {F.}~\bibnamefont {Liu}}, \bibinfo {author} {\bibfnamefont {C.}~\bibnamefont
  {Liu}}, \bibinfo {author} {\bibfnamefont {Y.}~\bibnamefont {Zhou}}, \bibinfo
  {author} {\bibfnamefont {N.}~\bibnamefont {Yang}}, \bibinfo {author}
  {\bibfnamefont {N.}~\bibnamefont {Zhong}}, \bibinfo {author} {\bibfnamefont
  {C.}~\bibnamefont {Duan}}, \ and\ \bibinfo {author} {\bibfnamefont
  {Y.}~\bibnamefont {Hao}},\ }\bibfield  {title} {\enquote {\bibinfo {title}
  {{Memory Behavior of an Al$_2$O$_3$ Gate Dielectric Non-Volatile Field-Effect
  Transistor}},}\ }\href {\doibase 10.1109/LED.2020.3010363} {\bibfield
  {journal} {\bibinfo  {journal} {IEEE Electron Device Letters}\ }\textbf
  {\bibinfo {volume} {41}},\ \bibinfo {pages} {1340--1343} (\bibinfo {year}
  {2020}{\natexlab{a}})}\BibitemShut {NoStop}%
\bibitem [{\citenamefont {Peng}\ \emph
  {et~al.}(2020{\natexlab{b}})\citenamefont {Peng}, \citenamefont {Han},
  \citenamefont {Liu}, \citenamefont {Xiao}, \citenamefont {Liu}, \citenamefont
  {Zhong}, \citenamefont {Duan}, \citenamefont {Feng}, \citenamefont {Dong},\
  and\ \citenamefont {Hao}}]{Peng2020}%
  \BibitemOpen
  \bibfield  {author} {\bibinfo {author} {\bibfnamefont {Y.}~\bibnamefont
  {Peng}}, \bibinfo {author} {\bibfnamefont {G.}~\bibnamefont {Han}}, \bibinfo
  {author} {\bibfnamefont {F.}~\bibnamefont {Liu}}, \bibinfo {author}
  {\bibfnamefont {W.}~\bibnamefont {Xiao}}, \bibinfo {author} {\bibfnamefont
  {Y.}~\bibnamefont {Liu}}, \bibinfo {author} {\bibfnamefont {N.}~\bibnamefont
  {Zhong}}, \bibinfo {author} {\bibfnamefont {C.}~\bibnamefont {Duan}},
  \bibinfo {author} {\bibfnamefont {Z.}~\bibnamefont {Feng}}, \bibinfo {author}
  {\bibfnamefont {H.}~\bibnamefont {Dong}}, \ and\ \bibinfo {author}
  {\bibfnamefont {Y.}~\bibnamefont {Hao}},\ }\bibfield  {title} {\enquote
  {\bibinfo {title} {{Ferroelectric-like Behavior Originating from Oxygen
  Vacancy Dipoles in Amorphous Film for Non-volatile Memory}},}\ }\href
  {\doibase 10.1186/s11671-020-03364-3} {\bibfield  {journal} {\bibinfo
  {journal} {Nanoscale Research Letters}\ }\textbf {\bibinfo {volume} {15}},\
  \bibinfo {pages} {0--5} (\bibinfo {year} {2020}{\natexlab{b}})}\BibitemShut
  {NoStop}%
\bibitem [{\citenamefont {Lim}\ \emph {et~al.}(2023)\citenamefont {Lim},
  \citenamefont {Lee}, \citenamefont {Lee}, \citenamefont {Choi}, \citenamefont
  {Jung}, \citenamefont {Baek},\ and\ \citenamefont {Jang}}]{Lim2023}%
  \BibitemOpen
  \bibfield  {author} {\bibinfo {author} {\bibfnamefont {T.}~\bibnamefont
  {Lim}}, \bibinfo {author} {\bibfnamefont {S.}~\bibnamefont {Lee}}, \bibinfo
  {author} {\bibfnamefont {J.}~\bibnamefont {Lee}}, \bibinfo {author}
  {\bibfnamefont {H.~J.}\ \bibnamefont {Choi}}, \bibinfo {author}
  {\bibfnamefont {B.}~\bibnamefont {Jung}}, \bibinfo {author} {\bibfnamefont
  {S.~H.}\ \bibnamefont {Baek}}, \ and\ \bibinfo {author} {\bibfnamefont
  {J.}~\bibnamefont {Jang}},\ }\bibfield  {title} {\enquote {\bibinfo {title}
  {{Artificial Synapse Based on Oxygen Vacancy Migration in Ferroelectric-Like
  C-Axis-Aligned Crystalline InGaSnO Semiconductor Thin-Film Transistors for
  Highly Integrated Neuromorphic Electronics}},}\ }\href {\doibase
  10.1002/adfm.202212367} {\bibfield  {journal} {\bibinfo  {journal} {Advanced
  Functional Materials}\ }\textbf {\bibinfo {volume} {33}},\ \bibinfo {pages}
  {1--12} (\bibinfo {year} {2023})}\BibitemShut {NoStop}%
\bibitem [{\citenamefont {Liu}\ \emph {et~al.}(2022)\citenamefont {Liu},
  \citenamefont {Li}, \citenamefont {Wang}, \citenamefont {Chen}, \citenamefont
  {Yu}, \citenamefont {Liu}, \citenamefont {Jin}, \citenamefont {Wang},
  \citenamefont {Hao},\ and\ \citenamefont {Han}}]{Liu2022}%
  \BibitemOpen
  \bibfield  {author} {\bibinfo {author} {\bibfnamefont {H.}~\bibnamefont
  {Liu}}, \bibinfo {author} {\bibfnamefont {J.}~\bibnamefont {Li}}, \bibinfo
  {author} {\bibfnamefont {G.}~\bibnamefont {Wang}}, \bibinfo {author}
  {\bibfnamefont {J.}~\bibnamefont {Chen}}, \bibinfo {author} {\bibfnamefont
  {X.}~\bibnamefont {Yu}}, \bibinfo {author} {\bibfnamefont {Y.}~\bibnamefont
  {Liu}}, \bibinfo {author} {\bibfnamefont {C.}~\bibnamefont {Jin}}, \bibinfo
  {author} {\bibfnamefont {S.}~\bibnamefont {Wang}}, \bibinfo {author}
  {\bibfnamefont {Y.}~\bibnamefont {Hao}}, \ and\ \bibinfo {author}
  {\bibfnamefont {G.}~\bibnamefont {Han}},\ }\bibfield  {title} {\enquote
  {\bibinfo {title} {{Analog Synapses Based on Nonvolatile FETs with Amorphous
  ZrO$_2$ Dielectric for Spiking Neural Network Applications}},}\ }\href
  {\doibase 10.1109/TED.2021.3139570} {\bibfield  {journal} {\bibinfo
  {journal} {IEEE Transactions on Electron Devices}\ }\textbf {\bibinfo
  {volume} {69}},\ \bibinfo {pages} {1028--1033} (\bibinfo {year}
  {2022})}\BibitemShut {NoStop}%
\bibitem [{\citenamefont {Zhang}\ \emph {et~al.}(2021)\citenamefont {Zhang},
  \citenamefont {Peng}, \citenamefont {Xiao}, \citenamefont {Liu},
  \citenamefont {Liu}, \citenamefont {Han},\ and\ \citenamefont
  {Hao}}]{Zhang2021}%
  \BibitemOpen
  \bibfield  {author} {\bibinfo {author} {\bibfnamefont {G.}~\bibnamefont
  {Zhang}}, \bibinfo {author} {\bibfnamefont {Y.}~\bibnamefont {Peng}},
  \bibinfo {author} {\bibfnamefont {W.}~\bibnamefont {Xiao}}, \bibinfo {author}
  {\bibfnamefont {F.}~\bibnamefont {Liu}}, \bibinfo {author} {\bibfnamefont
  {Y.}~\bibnamefont {Liu}}, \bibinfo {author} {\bibfnamefont {G.}~\bibnamefont
  {Han}}, \ and\ \bibinfo {author} {\bibfnamefont {Y.}~\bibnamefont {Hao}},\
  }\bibfield  {title} {\enquote {\bibinfo {title} {{Synaptic Plasticity in
  Novel Non-Volatile FET with Amorphous Gate Insulator Enabled by Oxygen
  Vacancy Related Dipoles}},}\ }\href {\doibase 10.1109/EDTM50988.2021.9420818}
  {\bibfield  {journal} {\bibinfo  {journal} {2021 5th IEEE Electron Devices
  Technology and Manufacturing Conference, EDTM 2021}\ ,\ \bibinfo {pages}
  {2021--2023}} (\bibinfo {year} {2021})}\BibitemShut {NoStop}%
\bibitem [{\citenamefont {Feng}\ \emph {et~al.}(2021)\citenamefont {Feng},
  \citenamefont {Peng}, \citenamefont {Shen}, \citenamefont {Li}, \citenamefont
  {Wang}, \citenamefont {Chen}, \citenamefont {Wang}, \citenamefont {Jing},
  \citenamefont {Lu}, \citenamefont {Wang}, \citenamefont {Cheng},
  \citenamefont {Cui}, \citenamefont {Dingsun}, \citenamefont {Han},
  \citenamefont {Liu},\ and\ \citenamefont {Dong}}]{Feng2021}%
  \BibitemOpen
  \bibfield  {author} {\bibinfo {author} {\bibfnamefont {Z.}~\bibnamefont
  {Feng}}, \bibinfo {author} {\bibfnamefont {Y.}~\bibnamefont {Peng}}, \bibinfo
  {author} {\bibfnamefont {Y.}~\bibnamefont {Shen}}, \bibinfo {author}
  {\bibfnamefont {Z.}~\bibnamefont {Li}}, \bibinfo {author} {\bibfnamefont
  {H.}~\bibnamefont {Wang}}, \bibinfo {author} {\bibfnamefont {X.}~\bibnamefont
  {Chen}}, \bibinfo {author} {\bibfnamefont {Y.}~\bibnamefont {Wang}}, \bibinfo
  {author} {\bibfnamefont {M.}~\bibnamefont {Jing}}, \bibinfo {author}
  {\bibfnamefont {F.}~\bibnamefont {Lu}}, \bibinfo {author} {\bibfnamefont
  {W.}~\bibnamefont {Wang}}, \bibinfo {author} {\bibfnamefont {Y.}~\bibnamefont
  {Cheng}}, \bibinfo {author} {\bibfnamefont {Y.}~\bibnamefont {Cui}}, \bibinfo
  {author} {\bibfnamefont {A.}~\bibnamefont {Dingsun}}, \bibinfo {author}
  {\bibfnamefont {G.}~\bibnamefont {Han}}, \bibinfo {author} {\bibfnamefont
  {H.}~\bibnamefont {Liu}}, \ and\ \bibinfo {author} {\bibfnamefont
  {H.}~\bibnamefont {Dong}},\ }\bibfield  {title} {\enquote {\bibinfo {title}
  {{Ferroelectric-Like Behavior in TaN/High-k/Si System Based on Amorphous
  Oxide}},}\ }\href {\doibase 10.1002/aelm.202100414} {\bibfield  {journal}
  {\bibinfo  {journal} {Advanced Electronic Materials}\ }\textbf {\bibinfo
  {volume} {7}} (\bibinfo {year} {2021}),\ 10.1002/aelm.202100414}\BibitemShut
  {NoStop}%
\bibitem [{\citenamefont {Mulaosmanovic}\ \emph {et~al.}(2017)\citenamefont
  {Mulaosmanovic}, \citenamefont {Ocker}, \citenamefont {Müller},
  \citenamefont {Noack}, \citenamefont {Müller}, \citenamefont {Polakowski},
  \citenamefont {Mikolajick},\ and\ \citenamefont {Slesazeck}}]{Mul2017}%
  \BibitemOpen
  \bibfield  {author} {\bibinfo {author} {\bibfnamefont {H.}~\bibnamefont
  {Mulaosmanovic}}, \bibinfo {author} {\bibfnamefont {J.}~\bibnamefont
  {Ocker}}, \bibinfo {author} {\bibfnamefont {S.}~\bibnamefont {Müller}},
  \bibinfo {author} {\bibfnamefont {M.}~\bibnamefont {Noack}}, \bibinfo
  {author} {\bibfnamefont {J.}~\bibnamefont {Müller}}, \bibinfo {author}
  {\bibfnamefont {P.}~\bibnamefont {Polakowski}}, \bibinfo {author}
  {\bibfnamefont {T.}~\bibnamefont {Mikolajick}}, \ and\ \bibinfo {author}
  {\bibfnamefont {S.}~\bibnamefont {Slesazeck}},\ }\bibfield  {title} {\enquote
  {\bibinfo {title} {Novel ferroelectric fet based synapse for neuromorphic
  systems},}\ }in\ \href {\doibase 10.23919/VLSIT.2017.7998165} {\emph
  {\bibinfo {booktitle} {2017 Symposium on VLSI Technology}}}\ (\bibinfo {year}
  {2017})\ pp.\ \bibinfo {pages} {T176--T177}\BibitemShut {NoStop}%
\bibitem [{\citenamefont {Peng}\ \emph {et~al.}(2019)\citenamefont {Peng},
  \citenamefont {Xu}, \citenamefont {Liu}, \citenamefont {Hao}, \citenamefont
  {Xiao}, \citenamefont {Han}, \citenamefont {Liu}, \citenamefont {Wu},
  \citenamefont {Wang}, \citenamefont {He}, \citenamefont {Yu},\ and\
  \citenamefont {Wang}}]{Peng2019}%
  \BibitemOpen
  \bibfield  {author} {\bibinfo {author} {\bibfnamefont {Y.}~\bibnamefont
  {Peng}}, \bibinfo {author} {\bibfnamefont {N.}~\bibnamefont {Xu}}, \bibinfo
  {author} {\bibfnamefont {T.~J.~K.}\ \bibnamefont {Liu}}, \bibinfo {author}
  {\bibfnamefont {Y.}~\bibnamefont {Hao}}, \bibinfo {author} {\bibfnamefont
  {W.}~\bibnamefont {Xiao}}, \bibinfo {author} {\bibfnamefont {G.}~\bibnamefont
  {Han}}, \bibinfo {author} {\bibfnamefont {Y.}~\bibnamefont {Liu}}, \bibinfo
  {author} {\bibfnamefont {J.}~\bibnamefont {Wu}}, \bibinfo {author}
  {\bibfnamefont {K.}~\bibnamefont {Wang}}, \bibinfo {author} {\bibfnamefont
  {Y.}~\bibnamefont {He}}, \bibinfo {author} {\bibfnamefont {Z.}~\bibnamefont
  {Yu}}, \ and\ \bibinfo {author} {\bibfnamefont {X.}~\bibnamefont {Wang}},\
  }\bibfield  {title} {\enquote {\bibinfo {title}
  {{Nanocrystal-Embedded-Insulator (NEI) Ferroelectric Field-Effect Transistor
  Featuring Low Operating Voltages and Improved Synaptic Behavior}},}\ }\href
  {\doibase 10.1109/LED.2019.2947086} {\bibfield  {journal} {\bibinfo
  {journal} {IEEE Electron Device Letters}\ }\textbf {\bibinfo {volume} {40}},\
  \bibinfo {pages} {1933--1936} (\bibinfo {year} {2019})}\BibitemShut {NoStop}%
\bibitem [{\citenamefont {Peng}\ \emph {et~al.}(2021)\citenamefont {Peng},
  \citenamefont {Zhang}, \citenamefont {Xiao}, \citenamefont {Liu},
  \citenamefont {Liu}, \citenamefont {Wang}, \citenamefont {Wang},
  \citenamefont {Yu}, \citenamefont {Han},\ and\ \citenamefont
  {Hao}}]{Peng2021}%
  \BibitemOpen
  \bibfield  {author} {\bibinfo {author} {\bibfnamefont {Y.}~\bibnamefont
  {Peng}}, \bibinfo {author} {\bibfnamefont {G.}~\bibnamefont {Zhang}},
  \bibinfo {author} {\bibfnamefont {W.}~\bibnamefont {Xiao}}, \bibinfo {author}
  {\bibfnamefont {F.}~\bibnamefont {Liu}}, \bibinfo {author} {\bibfnamefont
  {Y.}~\bibnamefont {Liu}}, \bibinfo {author} {\bibfnamefont {G.}~\bibnamefont
  {Wang}}, \bibinfo {author} {\bibfnamefont {S.}~\bibnamefont {Wang}}, \bibinfo
  {author} {\bibfnamefont {X.}~\bibnamefont {Yu}}, \bibinfo {author}
  {\bibfnamefont {G.}~\bibnamefont {Han}}, \ and\ \bibinfo {author}
  {\bibfnamefont {Y.}~\bibnamefont {Hao}},\ }\bibfield  {title} {\enquote
  {\bibinfo {title} {{Ferroelectric-Like Non-Volatile FET with Amorphous Gate
  Insulator for Supervised Learning Applications}},}\ }\href {\doibase
  10.1109/JEDS.2021.3120924} {\bibfield  {journal} {\bibinfo  {journal} {IEEE
  Journal of the Electron Devices Society}\ }\textbf {\bibinfo {volume} {9}},\
  \bibinfo {pages} {1145--1150} (\bibinfo {year} {2021})}\BibitemShut {NoStop}%
\bibitem [{\citenamefont {Peng}\ \emph {et~al.}(2022)\citenamefont {Peng},
  \citenamefont {Xiao}, \citenamefont {Zhang}, \citenamefont {Han},
  \citenamefont {Liu},\ and\ \citenamefont {Hao}}]{Peng2022}%
  \BibitemOpen
  \bibfield  {author} {\bibinfo {author} {\bibfnamefont {Y.}~\bibnamefont
  {Peng}}, \bibinfo {author} {\bibfnamefont {W.}~\bibnamefont {Xiao}}, \bibinfo
  {author} {\bibfnamefont {G.}~\bibnamefont {Zhang}}, \bibinfo {author}
  {\bibfnamefont {G.}~\bibnamefont {Han}}, \bibinfo {author} {\bibfnamefont
  {Y.}~\bibnamefont {Liu}}, \ and\ \bibinfo {author} {\bibfnamefont
  {Y.}~\bibnamefont {Hao}},\ }\bibfield  {title} {\enquote {\bibinfo {title}
  {{Synaptic Behaviors in Ferroelectric-Like Field-Effect Transistors with
  Ultrathin Amorphous HfO$_2$ Film}},}\ }\href {\doibase
  10.1186/s11671-022-03655-x} {\bibfield  {journal} {\bibinfo  {journal}
  {Nanoscale Research Letters}\ }\textbf {\bibinfo {volume} {17}} (\bibinfo
  {year} {2022}),\ 10.1186/s11671-022-03655-x}\BibitemShut {NoStop}%
\bibitem [{\citenamefont {Zheng}\ \emph {et~al.}(2021)\citenamefont {Zheng},
  \citenamefont {Zhou}, \citenamefont {Agarwal}, \citenamefont {Tang},
  \citenamefont {Zhang}, \citenamefont {Liu}, \citenamefont {Liu},
  \citenamefont {Han},\ and\ \citenamefont {Hao}}]{Zheng2021}%
  \BibitemOpen
  \bibfield  {author} {\bibinfo {author} {\bibfnamefont {S.}~\bibnamefont
  {Zheng}}, \bibinfo {author} {\bibfnamefont {J.}~\bibnamefont {Zhou}},
  \bibinfo {author} {\bibfnamefont {H.}~\bibnamefont {Agarwal}}, \bibinfo
  {author} {\bibfnamefont {J.}~\bibnamefont {Tang}}, \bibinfo {author}
  {\bibfnamefont {H.}~\bibnamefont {Zhang}}, \bibinfo {author} {\bibfnamefont
  {N.}~\bibnamefont {Liu}}, \bibinfo {author} {\bibfnamefont {Y.}~\bibnamefont
  {Liu}}, \bibinfo {author} {\bibfnamefont {G.}~\bibnamefont {Han}}, \ and\
  \bibinfo {author} {\bibfnamefont {Y.}~\bibnamefont {Hao}},\ }\bibfield
  {title} {\enquote {\bibinfo {title} {{Proposal of Ferroelectric Based
  Electrostatic Doping for Nanoscale Devices}},}\ }\href {\doibase
  10.1109/LED.2021.3063126} {\bibfield  {journal} {\bibinfo  {journal} {IEEE
  Electron Device Letters}\ }\textbf {\bibinfo {volume} {42}},\ \bibinfo
  {pages} {605--608} (\bibinfo {year} {2021})}\BibitemShut {NoStop}%
\bibitem [{\citenamefont {Chen}\ \emph {et~al.}(2022)\citenamefont {Chen},
  \citenamefont {Liu}, \citenamefont {Jin}, \citenamefont {Jia}, \citenamefont
  {Yu}, \citenamefont {Peng}, \citenamefont {Cheng}, \citenamefont {Chen},
  \citenamefont {Liu}, \citenamefont {Hao},\ and\ \citenamefont
  {Han}}]{Chen2022}%
  \BibitemOpen
  \bibfield  {author} {\bibinfo {author} {\bibfnamefont {J.}~\bibnamefont
  {Chen}}, \bibinfo {author} {\bibfnamefont {H.}~\bibnamefont {Liu}}, \bibinfo
  {author} {\bibfnamefont {C.}~\bibnamefont {Jin}}, \bibinfo {author}
  {\bibfnamefont {X.}~\bibnamefont {Jia}}, \bibinfo {author} {\bibfnamefont
  {X.}~\bibnamefont {Yu}}, \bibinfo {author} {\bibfnamefont {Y.}~\bibnamefont
  {Peng}}, \bibinfo {author} {\bibfnamefont {R.}~\bibnamefont {Cheng}},
  \bibinfo {author} {\bibfnamefont {B.}~\bibnamefont {Chen}}, \bibinfo {author}
  {\bibfnamefont {Y.}~\bibnamefont {Liu}}, \bibinfo {author} {\bibfnamefont
  {Y.}~\bibnamefont {Hao}}, \ and\ \bibinfo {author} {\bibfnamefont
  {G.}~\bibnamefont {Han}},\ }\bibfield  {title} {\enquote {\bibinfo {title}
  {{A Physics-Based Model for Mobile-Ionic Field-Effect Transistors with Steep
  Subthreshold Swing}},}\ }\href {\doibase 10.1109/JEDS.2022.3202928}
  {\bibfield  {journal} {\bibinfo  {journal} {IEEE Journal of the Electron
  Devices Society}\ } (\bibinfo {year} {2022}),\
  10.1109/JEDS.2022.3202928}\BibitemShut {NoStop}%
\bibitem [{\citenamefont {Wang}\ \emph {et~al.}(2015)\citenamefont {Wang},
  \citenamefont {Lin}, \citenamefont {Wang}, \citenamefont {Lin},\ and\
  \citenamefont {Hou}}]{Wang2015}%
  \BibitemOpen
  \bibfield  {author} {\bibinfo {author} {\bibfnamefont {Y.~F.}\ \bibnamefont
  {Wang}}, \bibinfo {author} {\bibfnamefont {Y.~C.}\ \bibnamefont {Lin}},
  \bibinfo {author} {\bibfnamefont {I.~T.}\ \bibnamefont {Wang}}, \bibinfo
  {author} {\bibfnamefont {T.~P.}\ \bibnamefont {Lin}}, \ and\ \bibinfo
  {author} {\bibfnamefont {T.~H.}\ \bibnamefont {Hou}},\ }\bibfield  {title}
  {\enquote {\bibinfo {title} {{Characterization and modeling of nonfilamentary
  Ta/TaOx/TiO$_2$/Ti analog synaptic device}},}\ }\href {\doibase
  10.1038/srep10150} {\bibfield  {journal} {\bibinfo  {journal} {Scientific
  Reports}\ }\textbf {\bibinfo {volume} {5}},\ \bibinfo {pages} {1--9}
  (\bibinfo {year} {2015})}\BibitemShut {NoStop}%
\bibitem [{\citenamefont {Kumar}\ \emph {et~al.}(2018)\citenamefont {Kumar},
  \citenamefont {{Balakrishna Pillai}}, \citenamefont {Song},\ and\
  \citenamefont {{De Souza}}}]{Kumar2018}%
  \BibitemOpen
  \bibfield  {author} {\bibinfo {author} {\bibfnamefont {A.}~\bibnamefont
  {Kumar}}, \bibinfo {author} {\bibfnamefont {P.}~\bibnamefont {{Balakrishna
  Pillai}}}, \bibinfo {author} {\bibfnamefont {X.}~\bibnamefont {Song}}, \ and\
  \bibinfo {author} {\bibfnamefont {M.~M.}\ \bibnamefont {{De Souza}}},\
  }\bibfield  {title} {\enquote {\bibinfo {title} {{Negative Capacitance beyond
  Ferroelectric Switches}},}\ }\href {\doibase 10.1021/acsami.8b05093}
  {\bibfield  {journal} {\bibinfo  {journal} {ACS Applied Materials and
  Interfaces}\ }\textbf {\bibinfo {volume} {10}},\ \bibinfo {pages}
  {19812--19819} (\bibinfo {year} {2018})}\BibitemShut {NoStop}%
\bibitem [{\citenamefont {Toral-Lopez}\ \emph {et~al.}(2022)\citenamefont
  {Toral-Lopez}, \citenamefont {Santos}, \citenamefont {Marin}, \citenamefont
  {Ruiz}, \citenamefont {Palacios},\ and\ \citenamefont
  {Godoy}}]{Toral-Lopez2022}%
  \BibitemOpen
  \bibfield  {author} {\bibinfo {author} {\bibfnamefont {A.}~\bibnamefont
  {Toral-Lopez}}, \bibinfo {author} {\bibfnamefont {H.}~\bibnamefont {Santos}},
  \bibinfo {author} {\bibfnamefont {E.~G.}\ \bibnamefont {Marin}}, \bibinfo
  {author} {\bibfnamefont {F.~G.}\ \bibnamefont {Ruiz}}, \bibinfo {author}
  {\bibfnamefont {J.~J.}\ \bibnamefont {Palacios}}, \ and\ \bibinfo {author}
  {\bibfnamefont {A.}~\bibnamefont {Godoy}},\ }\bibfield  {title} {\enquote
  {\bibinfo {title} {{Multi-scale modeling of 2D GaSe FETs with strained
  channels}},}\ }\href {\doibase 10.1088/1361-6528/ac3ce2} {\bibfield
  {journal} {\bibinfo  {journal} {Nanotechnology}\ }\textbf {\bibinfo {volume}
  {33}} (\bibinfo {year} {2022}),\ 10.1088/1361-6528/ac3ce2}\BibitemShut
  {NoStop}%
\bibitem [{\citenamefont {Scharfetter}\ and\ \citenamefont
  {Gummel}(1969)}]{SG1969}%
  \BibitemOpen
  \bibfield  {author} {\bibinfo {author} {\bibfnamefont {D.}~\bibnamefont
  {Scharfetter}}\ and\ \bibinfo {author} {\bibfnamefont {H.}~\bibnamefont
  {Gummel}},\ }\bibfield  {title} {\enquote {\bibinfo {title} {Large-signal
  analysis of a silicon read diode oscillator},}\ }\href {\doibase
  10.1109/T-ED.1969.16566} {\bibfield  {journal} {\bibinfo  {journal} {IEEE
  Transactions on Electron Devices}\ }\textbf {\bibinfo {volume} {16}},\
  \bibinfo {pages} {64--77} (\bibinfo {year} {1969})}\BibitemShut {NoStop}%
\bibitem [{\citenamefont {Foster}\ \emph {et~al.}(2002)\citenamefont {Foster},
  \citenamefont {Lopez~Gejo}, \citenamefont {Shluger},\ and\ \citenamefont
  {Nieminen}}]{Foster2002}%
  \BibitemOpen
  \bibfield  {author} {\bibinfo {author} {\bibfnamefont {A.~S.}\ \bibnamefont
  {Foster}}, \bibinfo {author} {\bibfnamefont {F.}~\bibnamefont {Lopez~Gejo}},
  \bibinfo {author} {\bibfnamefont {A.~L.}\ \bibnamefont {Shluger}}, \ and\
  \bibinfo {author} {\bibfnamefont {R.~M.}\ \bibnamefont {Nieminen}},\
  }\bibfield  {title} {\enquote {\bibinfo {title} {Vacancy and interstitial
  defects in hafnia},}\ }\href {\doibase 10.1103/PhysRevB.65.174117} {\bibfield
   {journal} {\bibinfo  {journal} {Phys. Rev. B}\ }\textbf {\bibinfo {volume}
  {65}},\ \bibinfo {pages} {174117} (\bibinfo {year} {2002})}\BibitemShut
  {NoStop}%
\bibitem [{\citenamefont {S.M.~Sze}\ and\ \citenamefont {Ng}(2012)}]{Sze2012}%
  \BibitemOpen
  \bibfield  {author} {\bibinfo {author} {\bibfnamefont {Y.~L.}\ \bibnamefont
  {S.M.~Sze}}\ and\ \bibinfo {author} {\bibfnamefont {K.}~\bibnamefont {Ng}},\
  }\href@noop {} {\emph {\bibinfo {title} {{Physics of semiconductor
  devices}}}}\ (\bibinfo  {publisher} {Wiley},\ \bibinfo {year}
  {2012})\BibitemShut {NoStop}%
\bibitem [{\citenamefont {Ceresoli}\ and\ \citenamefont
  {Vanderbilt}(2006)}]{Ceresoli2006}%
  \BibitemOpen
  \bibfield  {author} {\bibinfo {author} {\bibfnamefont {D.}~\bibnamefont
  {Ceresoli}}\ and\ \bibinfo {author} {\bibfnamefont {D.}~\bibnamefont
  {Vanderbilt}},\ }\bibfield  {title} {\enquote {\bibinfo {title} {{Structural
  and dielectric properties of amorphous ZrO$_2$ and HfO$_2$}},}\ }\href
  {\doibase 10.1103/PhysRevB.74.125108} {\bibfield  {journal} {\bibinfo
  {journal} {Physical Review B - Condensed Matter and Materials Physics}\
  }\textbf {\bibinfo {volume} {74}},\ \bibinfo {pages} {2--7} (\bibinfo {year}
  {2006})}\BibitemShut {NoStop}%
\bibitem [{\citenamefont {S.~Zafar}\ and\ \citenamefont
  {Gupta}(2010)}]{Zafar2010}%
  \BibitemOpen
  \bibfield  {author} {\bibinfo {author} {\bibfnamefont {L.~F.~E.}\
  \bibnamefont {S.~Zafar}, \bibfnamefont {H.~Jagannathan}}\ and\ \bibinfo
  {author} {\bibfnamefont {D.}~\bibnamefont {Gupta}},\ }\bibfield  {title}
  {\enquote {\bibinfo {title} {{Oxygen vacancy mobility and diffusion
  coefficient determined from current measurements in SiO$_2$/HfO$_2$/TiN
  stacks}},}\ }\href@noop {} {\bibfield  {journal} {\bibinfo  {journal}
  {Proceedings of the International Conference on Solid State Devices and
  Materials}\ ,\ \bibinfo {pages} {669}} (\bibinfo {year} {2010})}\BibitemShut
  {NoStop}%
\bibitem [{\citenamefont {Marchewka}\ \emph {et~al.}(2016)\citenamefont
  {Marchewka}, \citenamefont {Roesgen}, \citenamefont {Skaja}, \citenamefont
  {Du}, \citenamefont {Jia}, \citenamefont {Mayer}, \citenamefont {Rana},
  \citenamefont {Waser},\ and\ \citenamefont {Menzel}}]{Marchewka2016}%
  \BibitemOpen
  \bibfield  {author} {\bibinfo {author} {\bibfnamefont {A.}~\bibnamefont
  {Marchewka}}, \bibinfo {author} {\bibfnamefont {B.}~\bibnamefont {Roesgen}},
  \bibinfo {author} {\bibfnamefont {K.}~\bibnamefont {Skaja}}, \bibinfo
  {author} {\bibfnamefont {H.}~\bibnamefont {Du}}, \bibinfo {author}
  {\bibfnamefont {C.~L.}\ \bibnamefont {Jia}}, \bibinfo {author} {\bibfnamefont
  {J.}~\bibnamefont {Mayer}}, \bibinfo {author} {\bibfnamefont
  {V.}~\bibnamefont {Rana}}, \bibinfo {author} {\bibfnamefont {R.}~\bibnamefont
  {Waser}}, \ and\ \bibinfo {author} {\bibfnamefont {S.}~\bibnamefont
  {Menzel}},\ }\bibfield  {title} {\enquote {\bibinfo {title} {{Nanoionic
  Resistive Switching Memories: On the Physical Nature of the Dynamic Reset
  Process}},}\ }\href {\doibase 10.1002/aelm.201500233} {\bibfield  {journal}
  {\bibinfo  {journal} {Advanced Electronic Materials}\ }\textbf {\bibinfo
  {volume} {2}} (\bibinfo {year} {2016}),\ 10.1002/aelm.201500233}\BibitemShut
  {NoStop}%
\bibitem [{\citenamefont {Daus}\ \emph {et~al.}(2017)\citenamefont {Daus},
  \citenamefont {Lenarczyk}, \citenamefont {Petti}, \citenamefont
  {M{\"{u}}nzenrieder}, \citenamefont {Knobelspies}, \citenamefont
  {Cantarella}, \citenamefont {Vogt}, \citenamefont {Salvatore}, \citenamefont
  {Luisier},\ and\ \citenamefont {Tr{\"{o}}ster}}]{Daus2017}%
  \BibitemOpen
  \bibfield  {author} {\bibinfo {author} {\bibfnamefont {A.}~\bibnamefont
  {Daus}}, \bibinfo {author} {\bibfnamefont {P.}~\bibnamefont {Lenarczyk}},
  \bibinfo {author} {\bibfnamefont {L.}~\bibnamefont {Petti}}, \bibinfo
  {author} {\bibfnamefont {N.}~\bibnamefont {M{\"{u}}nzenrieder}}, \bibinfo
  {author} {\bibfnamefont {S.}~\bibnamefont {Knobelspies}}, \bibinfo {author}
  {\bibfnamefont {G.}~\bibnamefont {Cantarella}}, \bibinfo {author}
  {\bibfnamefont {C.}~\bibnamefont {Vogt}}, \bibinfo {author} {\bibfnamefont
  {G.~A.}\ \bibnamefont {Salvatore}}, \bibinfo {author} {\bibfnamefont
  {M.}~\bibnamefont {Luisier}}, \ and\ \bibinfo {author} {\bibfnamefont
  {G.}~\bibnamefont {Tr{\"{o}}ster}},\ }\bibfield  {title} {\enquote {\bibinfo
  {title} {{Ferroelectric-Like Charge Trapping Thin-Film Transistors and Their
  Evaluation as Memories and Synaptic Devices}},}\ }\href {\doibase
  10.1002/aelm.201700309} {\bibfield  {journal} {\bibinfo  {journal} {Advanced
  Electronic Materials}\ }\textbf {\bibinfo {volume} {3}},\ \bibinfo {pages}
  {1--9} (\bibinfo {year} {2017})}\BibitemShut {NoStop}%
\bibitem [{\citenamefont {Farronato}\ \emph {et~al.}(2022)\citenamefont
  {Farronato}, \citenamefont {Melegari}, \citenamefont {Ricci}, \citenamefont
  {Hashemkhani}, \citenamefont {Bricalli},\ and\ \citenamefont
  {Ielmini}}]{Farronato2022}%
  \BibitemOpen
  \bibfield  {author} {\bibinfo {author} {\bibfnamefont {M.}~\bibnamefont
  {Farronato}}, \bibinfo {author} {\bibfnamefont {M.}~\bibnamefont {Melegari}},
  \bibinfo {author} {\bibfnamefont {S.}~\bibnamefont {Ricci}}, \bibinfo
  {author} {\bibfnamefont {S.}~\bibnamefont {Hashemkhani}}, \bibinfo {author}
  {\bibfnamefont {A.}~\bibnamefont {Bricalli}}, \ and\ \bibinfo {author}
  {\bibfnamefont {D.}~\bibnamefont {Ielmini}},\ }\bibfield  {title} {\enquote
  {\bibinfo {title} {{Memtransistor Devices Based on MoS$_2$ Multilayers with
  Volatile Switching due to Ag Cation Migration}},}\ }\href {\doibase
  10.1002/aelm.202101161} {\bibfield  {journal} {\bibinfo  {journal} {Advanced
  Electronic Materials}\ }\textbf {\bibinfo {volume} {8}},\ \bibinfo {pages}
  {1--7} (\bibinfo {year} {2022})}\BibitemShut {NoStop}%
\bibitem [{\citenamefont {Chekol}\ \emph {et~al.}(2022)\citenamefont {Chekol},
  \citenamefont {Menzel}, \citenamefont {Ahmad}, \citenamefont {Waser},\ and\
  \citenamefont {Hoffmann-Eifert}}]{Chekol2022}%
  \BibitemOpen
  \bibfield  {author} {\bibinfo {author} {\bibfnamefont {S.~A.}\ \bibnamefont
  {Chekol}}, \bibinfo {author} {\bibfnamefont {S.}~\bibnamefont {Menzel}},
  \bibinfo {author} {\bibfnamefont {R.~W.}\ \bibnamefont {Ahmad}}, \bibinfo
  {author} {\bibfnamefont {R.}~\bibnamefont {Waser}}, \ and\ \bibinfo {author}
  {\bibfnamefont {S.}~\bibnamefont {Hoffmann-Eifert}},\ }\bibfield  {title}
  {\enquote {\bibinfo {title} {Effect of the threshold kinetics on the filament
  relaxation behavior of ag-based diffusive memristors},}\ }\href {\doibase
  https://doi.org/10.1002/adfm.202111242} {\bibfield  {journal} {\bibinfo
  {journal} {Advanced Functional Materials}\ }\textbf {\bibinfo {volume}
  {32}},\ \bibinfo {pages} {2111242} (\bibinfo {year} {2022})}\BibitemShut
  {NoStop}%
\end{thebibliography}%

\end{document}